\documentclass[twocolumn]{aastex61}

\hypersetup{linkcolor=blue,citecolor=blue,filecolor=cyan,urlcolor=magenta}

\usepackage{natbib}

\newcommand\aastex{AAS\TeX}

\graphicspath{{./img/}}
\received{01 Feb 2018}
\revised{14 March 2018}
\accepted{23 March 2018}

\shorttitle{\aastex\ Upper limits on central massive black holes in two UCDs in Cen\,A}
\shortauthors{Voggel et al.}

\begin{document}
 
\title{Upper limits on the presence of central massive black holes in two ultra-compact dwarf galaxies in Centaurus\,A}
\correspondingauthor{Karina Voggel}
\email{kvoggel@astro.utah.edu}

\author{Karina T. Voggel} 
\affiliation{University of Utah. Department of Physics \& Astronomy, 115 S 1400 E, Salt Lake City, UT 84105} 
\author{Anil C. Seth}
\affiliation{University of Utah. Department of Physics \& Astronomy, 115 S 1400 E, Salt Lake City, UT 84105}
\author{Nadine Neumayer}
\affiliation{Max-Planck Institut f\"ur Astronomie, K\"onigstuhl 17, 69117 Heidelberg, Germany} 
\author{Steffen Mieske}
\affiliation{European Southern Observatory,Alonso de Cordova 3107, Vitacura, Santiago, Chile}
\author{Igor Chilingarian}
\affiliation{Smithsonian Astrophysical Observatory, 60 Garden St. MS09, 02138 Cambridge, MA, USA}
\affiliation{Sternberg Astronomical Institute, M.V. Lomonosov Moscow State University, 13 Universitetsky prospect, 119992 Moscow, Russia}
\author{Christopher Ahn}
\affiliation{University of Utah. Department of Physics \& Astronomy, 115 S 1400 E, Salt Lake City, UT 84105}
\author{Holger Baumgardt}
\affiliation{School of Mathematics and Physics, University of Queensland, St Lucia, QLD 4072, Australia}
\author{Michael Hilker}
\affiliation{European Southern Observatory, Karl-Schwarzschild-Str. 2, 85748 Garching bei M\"unchen, Germany}
\author{Dieu D. Nguyen}
\affiliation{University of Utah. Department of Physics \& Astronomy, 115 S 1400 E, Salt Lake City, UT 84105}
\author{Aaron J. Romanowsky}
\affiliation{Department of Physics and Astronomy, San JosŽ State University, San Jose, CA 95192, USA}
\affiliation{University of California Observatories, 1156 High St., Santa Cruz, CA 95064, USA}
\author{Jonelle L. Walsh}
\affiliation{George P. and Cynthia Woods Mitchell Institute for Fundamental Physics and Astronomy, Department of Physics and Astronomy, Texas A\&M University, College Station, TX 77843, USA}
\author{Mark den Brok}
\affiliation{Institute for Astronomy, Department of Physics, ETH Zurich, CH-8093 Z\"urich, Switzerland}
\author{Jay Strader}
\affiliation{Center for Data Intensive and Time Domain Astronomy, Department of Physics and Astronomy, Michigan State University, 567 Wilson
Road, East Lansing, MI 48824, USA}





\begin{abstract}

The recent discovery of massive black holes (BHs) in the centers of high-mass ultra compact dwarf galaxies (UCDs) suggests that at least some are the stripped nuclear star clusters of dwarf galaxies. We present the first study that investigates whether such massive BHs, and therefore stripped nuclei, also exist in low-mass ($M<10^{7}M_{\odot}$) UCDs. 
We constrain the BH masses of two UCDs located in Centaurus\,A (UCD\,320 and UCD\,330) using Jeans modeling of the resolved stellar kinematics from adaptive optics VLT/SINFONI data. No massive BHs are found in either UCD.
We find a $3\,\sigma$ upper limit on the central BH mass in UCD\,330 of $M_{\bullet}<1.0\times10^{5}M_{\odot}$, which corresponds to 1.7\% of the total mass. This excludes a high mass fraction BH and would only allow a low-mass BHs similar to those claimed to be detected in Local Group GCs. For UCD\,320, poorer data quality results in a less constraining $3\,\sigma$ upper limit of $M_{\bullet}<1\times10^{6}M_{\odot}$, which is equal to 37.7\% of the total mass.
The dynamical $M/L$ of UCD\,320 and UCD\,330 are not inflated compared to predictions from stellar population models. The non-detection of BHs in these low-mass UCDs is consistent with the idea that elevated dynamical $M/L$s do indicate the presence of a substantial BH.
Despite not detecting massive BHs, these systems could still be stripped nuclei.
The strong rotation ($v/\sigma$ of 0.3 to 0.4) in both UCDs and the two-component light profile in UCD\,330 support the idea that these UCDs may be stripped nuclei of low-mass galaxies where the BH occupation fraction is not yet known.

\end{abstract}

\keywords{galaxies: kinematics and dynamics  - galaxies: dwarfs - galaxies: nuclei - galaxies: star clusters }



\section{Introduction} 
\label{sec:intro}

Ultra-compact dwarf systems (UCDs) are among the densest stellar objects in the universe and with their almost spherical appearances, they resemble globular clusters (GCs) (\citealt{Minniti1998, Hilker1999, Drinkwater2000}). A common definition of UCDs is that they have to be more massive than $\omega$\,Cen ($M> 2\times 10^{6}M_{\odot}$). But there is no clear physical property that separates UCDs from GCs. Yet when compared to dwarf galaxies they are much smaller and have higher stellar densities at the same luminosity (\citealt{Misgeld2011, Norris2014}). It is still under debate how these ''intermediate" objects formed.

One proposed formation channel for UCDs is that they formed as genuine massive globular clusters (\citealt{Murray2009, Mieske2004, Mieske2012}), during intense starbursts or mergers that have high enough star-formation rates to produce such massive clusters (\citealt{Schulz2015, Renaud2015}). Young clusters in the UCD mass range have been observed in nearby merger remnants, with virial masses up to $8\times10^{7}M_{\odot}$ (\citealt{Maraston2004, Bastian2006}).
A second formation mechanism is that UCDs might be the stripped nuclear star cluster of a parent galaxy that was accreted onto a larger galaxy or galaxy cluster (\citealt{Bekki2003, Drinkwater2003, Pfeffer2013}). 

There is evidence that supports the notion that both formation channels contribute to the population of UCDs we observe (\citealt{Hilker2006, DaRocha2011, Brodie2011, Norris2011}). However, it is unclear so far what fraction of UCDs was formed as genuine GCs and how many of them are former galaxy nuclei. Related questions are whether the contribution of UCD formation channels changes with UCD mass and environment and depends on the galaxy cluster they reside in.

The number of stripped nuclei in Fornax and Virgo cluster environment was predicted using the Millennium\,II simulation and the associated semi-analytic model \cite{Pfeffer2014, Pfeffer2016}. Its estimated that above masses of $10^{7}M_{\odot}$ stripped nuclei make up 40\% of all objects in the Fornax cluster and the most massive globular cluster would have a mass of $2\times10^{7}M_{\odot}$. The fraction of stripped nuclei drops significantly to 2.5\% between $10^{6}-10^{7}M_{\odot}$. Overall the combined mass function of simulated stripped nuclei and GCs agrees well with observations, indicating that UCDs are indeed a mix between GCs and stripped nuclei.

Quantifying the number of stripped nuclei in a galaxy cluster would provide a new way to infer its past merger history. Stripped nuclei UCDs could then provide a useful anchor point for simulations that predict the number of tidally disrupted dark matter halos in a galaxy cluster. 
 
There are three main ways to identify UCDs as stripped nuclei: 1.) detecting the remnant tails and extended low-surface brightness envelopes caused by the tidal stripping process, 2.) determining whether a UCDs star-formation history is extended and 3.) measuring whether they host a super-massive black hole (SMBH) in their centers, which are common in nuclei of galaxies. 

Tidal tails and envelopes around UCDs are expected when a galaxy is in process of being stripped, but these typically have short lifetimes. Such features were detected around UCDs in the Fornax and the Perseus cluster (\citealt{Voggel2016a, Wittmann2016}). A tidal stream of 1.5\,kpc was recently found around a newly discovered very massive ($M=4.2\times10^{8}M_{\odot}$) UCD in NGC\,7727 \citep{Schweizer2018}. In addition, a UCD of the size of $\omega\,Cen$ was discovered embedded in a stellar stream around NGC\,3628 \citep{Jennings2015}. 

An extended star formation history that extends over several Gyr has been found in NGC 4546-UCD1 (\citealt{Norris2015}). This long star formation timescale is similar to what is observed in galaxy nuclei (e.g. \citealt{Rossa2006, Seth2006, Walcher2006}). In contrast GCs have usually very short ($<$1Gyr) star formation histories. In the Milky Way, two massive clusters have extended star formation histories. The first is M54, the nucleus of the partially stripped Sgr dwarf galaxy \citep{Siegel2007, Carretta2010}, and the other is Omega Cen \citep{Hilker2004}, which is widely thought to be a stripped nucleus.

If UCDs are stripped nuclei then we expect super-massive black holes in their centers, similar to those observed in the nuclear star clusters of galaxies \citep{Seth2008b, Graham2009}. Due to the large mass of a SMBH it causes a distinctive central rise of the velocity dispersion that is detectable in bright UCDs using adaptive optics combined with integral field spectroscopy. Such BH mass measurements have been carried out in four high-mass UCDs ($>10^{7}M_{\odot}$) and there is strong observational evidence from dynamical modeling that they all host super-massive black holes (SMBHs) that make up $\sim15\%$ of their total mass \citep[][Afanasiev et al., {\em in prep}]{Seth2014, Ahn2017}. 
The higher than expected velocity dispersions of these massive UCDs also provides indirect evidence for a high fraction of SMBHs and thus  suggests a high fraction of former galaxy nuclei among high-mass UCDs.

At the low-mass end, there is evidence that both M54 and $\omega\,Cen$ have a massive BH in their centers and are thus stripped nuclei. In M\,54 as BH mass of $1\times10^{4}M_{\odot}$ was suggested (\citealt{Ibata2009}), and a $4.0-4.7\times 10^{4}M_{\odot}$ black hole is suggested in the center of $\omega$\,Cen (\citealt{Noyola2010, Baumgardt2017}). The central dispersion increase of such intermediate mass BHs could also be explained with significant radial anisotropy without an IMBH (\citealt{vanderMarel2010, Zocchi2017}).

If UCDs are the remnant nuclear star clusters (NSCs) of a stripped galaxy, then their masses directly trace the mass of the progenitor host galaxy via the NSC--host galaxy mass relation \citep{Ferrarese2006}. However, this relation has a significant scatter meaning that galaxies of the same mass can have nuclei masses that vary by two orders of magnitude \citep{Georgiev2016}. In the scenario where UCDs ($M>2\times10^{6}M_{\odot}$) are the stripped nuclei of former more massive galaxies, they will trace the merging of progenitor galaxies with stellar masses of $5\times10^{8}M_{\odot}<M<10^{11}M_{\odot}$ assuming the nuclei--galaxy mass correlation (\citealt{Georgiev2016}). The high metallicities of UCDs with confirmed SMBHs are consistent with them being nuclei that follow the mass--metallicity relation of their larger parent galaxy (\citealt{Tremonti2004}).

Resolved kinematic studies of UCDs are only feasible for the brightest UCDs and thus our existing sample is strongly biased towards more massive UCDs ($>10^{7}M_{\odot}$), while in fact there are many more UCDs at lower masses. There is no measurement of the presence of SMBHs in lower mass UCDs yet, and the incidence of genuine nuclei is entirely unknown for low-mass UCDs. If stripped nuclei exist among low-mass UCDs they most likely originate from low-mass ($\sim1\times10^{9}M_{\odot}$) parent galaxies. For this mass-range, the BH demographics are not well known, but even these low-mass nuclei appear to host BHs \citep{Miller2015, Nguyen2017}. To provide a first look inside lower mass UCDs we target two UCDs below $10^{7}M_{\odot}$ in this work to explore whether those also host SMBHs in their centers. Due to their lower brightness, the required AO observations are only feasible for UCDs that are closer than the Fornax or Virgo clusters. 

We chose to target two Centaurus\,A UCDs (UCD\,320 and UCD\,330, also named HGHH92-C21 and HGHH92-C23 respectively; see \cite{Taylor2010, Rejkuba2007} for reference) that are both more massive than $\omega$\,Cen. We show below that they have masses of $2.8\times10^{6}M_{\odot}$ and $6.1\times10^{6}M_{\odot}$ for UCD\,320 and UCD\,330 respectively and their other properties are summarized in Table \ref{tab:UCDs}.

Both objects have dynamical mass-to-light ratios ($M/L_{\rm dyn}$) that are higher than what is expected from stellar population predictions ($M/L_{\rm pop}$). Based on the SMBHs found in massive UCDs, this inflated M/L may be a sign of massive BHs in these systems \citep{Ahn2017}.
The enhancement for UCD\,320 was $\Psi_{330}=\frac{M/L_{\rm dyn}}{M/L_{\rm pop}}=2.28$ and $\Psi_{320}=\frac{M/L_{\rm dyn}}{M/L_{\rm pop}}=2.5$ for UCD\,330.

  \begin{deluxetable}{lccc}
\label{tab:UCDs}
\tablecaption{Literature values for UCD\,330 and UCD\,320}
\tablehead{\colhead{Name} & \colhead{UCD\,330} & \colhead{UCD\,320} & \colhead{Reference} } 
\startdata
R.A. & 13:25:54.3 & 13:25:52.7 & Taylor+2010 \\ 
Dec. & $-$42:59:25.4  & $-$43:05:46.6 & Taylor+2010  \\
$M_{\rm V}$ [mag]& $-$11.66 & $-$10.39 & Rejkuba+2007 \\
$[M/H]$ & $-$0.36$\pm$0.14 & $-$0.85$\pm$0.14  & Beasley+2008 \\
$r_{\rm eff}$ [pc] & 3.25$\pm$0.13 & 6.83$\pm$0.10 & Taylor+2010 \\
$\sigma_{\rm v}$ [$\rm km\,s^{-1}$] & 41.5$\pm$3.7 & 20.0$\pm$1.4 & Taylor+2010 \\
$\sigma_{\rm v}$ [$\rm km\,s^{-1}$] & 30.5$\pm$0.2 & 19.0$\pm$0.1 & Rejkuba+2007 \\
$R_{\rm gc}$ [kpc] & 5.8 & 7.3 & Rejkuba+2007 \\
\enddata
\end{deluxetable}

These observations are part of an adaptive optics campaign that uses the VLT/SINFONI (PI: Mieske) and for the UCDs in the northern hemisphere with Gemini/NIFS (PI: Seth). Both UCDs have been first noted in \cite{Harris1992} and their integrated velocity dispersion was measured in \cite{Rejkuba2007}. The data of \cite{Rejkuba2007} were reanalyzed by \cite{Taylor2010}.

In the paper we adopt a distance modulus of $m-M = 27.91$ to Cen\,A (\citealt{Harris2010}) and an extinction value of $A_{\rm V}=0.31\,mag$.

This paper is organised in the following way:  In Section \ref{sec:data} we present our data and how they were analyzed. 
In Section \ref{sec:kinmethod} we present our methods for measuring the kinematics and the mass and surface-brightness profile of the UCD and the set-up of the Jeans Anisotropic Models (JAM). In Section \ref{sec:results} we present our results from the kinematic measurements and in Section \ref{sec:dyn_models} the results from the dynamical modeling. In Section \ref{sec:discussion} we discuss our findings and in Section \ref{sec:conclusion} we present the conclusion. The appendix contains the tables with the luminosity profiles.

\section{Data} 
\label{sec:data}
\subsection{SINFONI Observations}

UCD\,320 and UCD\,330 were observed with SINFONI \citep{Eisenhauer2003}
on UT4 of the VLT, under ESO ID Nr.095.B-0451(A) (PI: Mieske).
SINFONI is a near-infrared integral field spectrograph with adaptive optics capabilities. All our observations were carried out in the $K$-band (1.95-2.45$\mu$m) with a pixel scale of 50$\times$100mas, a field-of-view of 3$\arcsec\times3\arcsec$ and a spectral resolution of R$\sim$4000.

For UCD\,330 in total 21 exposures of 600\,s were combined into the final cube. The observations were carried out on the nights of the 15, 18, 21 and 24th of June 2015.

For UCD\,320 we observed 28 exposures of 600\,s and had to discard 6 of those due to low quality where the adaptive optics loop was not stable. For the final cube we combine 22 exposures.

The data were reduced using version 3.12.3 of the esorex command-line software and version 1.8.2 of the SINFONI instrument pipeline. 
We correct each individual observation with the dark exposure and apply the pipeline recipes that correct linearity and distortion. We then divide by the flat field, apply a wavelength correction and correct for the telluric absorption features. The sky was subtracted using the two offset sky exposures taken in each observing block in a O-S-O-S-O sequence, with offsets of $4\arcsec$ and $7\arcsec$ from the center of the UCD. The individual cubes were dithered in such a way that the object falls half of the total exposure time onto the lower right part of the detector and the other half on the upper left part. Additionally, a dither of a few pixels was applied between successive exposures at both positions, to ensure that the UCDs do not fall into the same area of the detector each time. This ensures that systematic detector effects are minimized, and that unique sky pixels are subtracted from each dither position. The individual cubes were combined using our own routine that centers on each UCD and co-adds them so they are aligned.

Despite the sky subtraction, the reduced cubes still had significant background flux left in the spectra. This residual background was uniform in spectral distribution across the chip, but had neither the spectrum expected for a stellar source or sky emission. We suspect the background is due to scattered light, similar to backgrounds seen in comparable SINFONI data \citep{Nguyen2017}. We estimate the background spectrum by using spatial pixels furthest from the center of the UCD, average these pixels using sigma clipping, and subtract this background spectrum from each spatial pixel in the cube. This background correction resulted in significantly improved kinematic fits, but has the consequence of introducing uncertainty to our PSF (see below).

The intrinsic dispersion of SINFONI varies for each row of the $64\times64$ pixel detector and thus we need to obtain accurate instrumental dispersion for each row separately. To achieve this we use five strong OH sky lines with small wavelength separations between the doublets from the sky cubes. For each line we subtract the continuum, normalize the flux in each line and then sum over all lines and take their median. Thus we use the empirically determined median line shape of each row as the instrumental dispersion of SINFONI. The line-spread function (LSF) of SINFONI varies significantly from row to row with FWHMs ranging from 5.7\,\AA \, up to 8.5\,\AA. We then dither the LSF cube of the SINFONI field-of-view in the same manner that our observations were dithered to create a final combined LSF cube. 

We derive the spatial point-spread function (PSF) of the SINFONI adaptive optics data by convolving the \textit{HST} images (see Sec. \ref{hstdata}) of the UCDs with a model PSF and comparing it to the collapsed image from the SINFONI cubes. For the model PSF we use a double Gaussian functional form. The double Gaussian model parameters are varied until a best-fit convolved \textit{HST} image is found that is closest to the observed SINFONI data.

The additional background subtraction we applied to the SINFONI cube reduced the light in the outskirts relative to the true distribution, potentially impacting our PSF measurement. To measure the accurate surface brightness profile, we needed to quantify what fraction of the signal in the outskirts comes from the UCD (and potentially galaxy) light. We extract 14 background aperture spectra (using four pixel apertures) at large radii ($>1,8\arcsec$) from UCD\,330. We then compare the $2.3\,\rm\mu m$ CO bandhead equivalent widths (EWs) of the background spectra to one from the center of the UCD.
No clear CO lines are visible in the background spectra, and from the equivalent width comparison we constrain that the contribution of a UCD-like spectrum to the background is $10.5\pm 7.7 \%$ at a radius of $\sim$2.3''.  Because the background spectra have strong structure in wavelength (i.e.~it looks like an emission line spectrum), one way to estimate the true surface brightness of the UCD is to make an image out of a region without strong emission in the background spectra. For determining the PSF, we therefore create an image by collapsing the data cube over wavelengths from $2.26-2.36\rm \mu m$, and then remove $89.5\pm7.7\,\%$ of the background level at 2.3'' to ensure that the scattered light from the UCD remains. The final PSF for UCD\,330 has a inner Gaussian width of $0.07\arcsec$ containing $72.9\,\%$ of the total luminosity, and an outer component of $1.15\arcsec$ that contains $27.1\%$ of the light. Considering the uncertainties in the kinematic PSF light contribution, we determine the following values for the outer Gaussian component: $r=0.97\arcsec$ with a $20.5\%$ light fraction as a lower limit, and $r=1.34\arcsec$ with a $35\%$ light fraction as the upper limit. The size of the inner component remained the same in both fits.

For UCD\,320 this method results in the inner Gaussian having a FWHM of $0.16\arcsec$ containing $59.8\,\%$ of the light and the outer component has a size of $0.85\arcsec$ and a light fraction of $40.2\,\%$. Using the equivalent width method, we find a UCD light contribution of $9\pm6\%$. The change in UCD light contribution varied the light fraction in the outer gaussian only by a small amount, with $40^{+3}_{-2}\%$. The sizes of the inner and outer gaussian were essentially unchanged.

\subsection{HST Data}
\label{hstdata}
High resolution imaging data from \textit{HST} was available on the Hubble Legacy Archive \footnote{\url{https://hla.stsci.edu/}} for both UCDs. The available imaging data were taken with the Wide Field Camera (WFC) on the Advanced Camera for Surveys (ACS) using the F606W filter. The combination of ACS/WFC provides a spatial resolution of $0.05\,\arcsec/\rm pix$. UCD\,320 and UCD\,330 were observed as part of \textit{HST} Proposal 10597(PI: Jordan) that targeted the structural parameters of GCs around Cen\,A. The total observing time was 2158\,s. We note that due to the single band of data, we cannot study color variations within the UCD or variations that would affect our assumption of a constant M/L, but as shown in \cite{Ahn2017}, these variations even if present have minimal effects on the dynamical models.

   \begin{figure}
   \centering
   \includegraphics[width=\hsize]{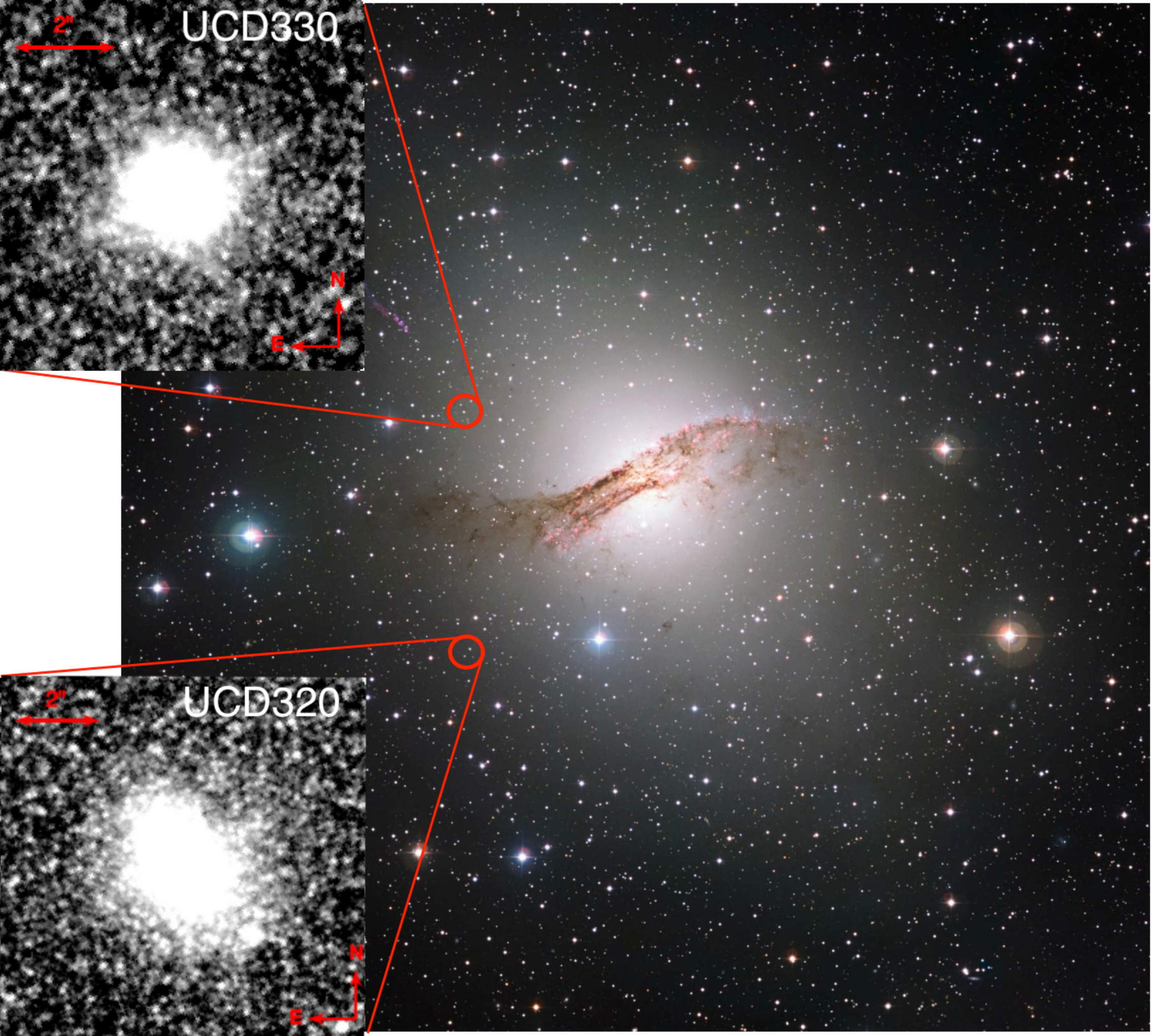}
      \caption{A cut-out of the \textit{HST} imaging of UCD\,330 (top panel) and UCD\,320 (bottom panel) are shown with respect to their position within Cen\,A. UCD\,330 lies at a distance of 5.8\,kpc to the center of Cen\,A and UCD\,320 is located at 7.3\,kpc. The image of Cen\,A is a publicly available composite image taken from ESO (\url{https://www.eso.org/public/images/eso1221a/}). }
         \label{fig:overview}
   \end{figure}  
The images are available in their fully reduced form from the \textit{HST} archive. We use them to analyze the surface-brightness profile of the UCDs and determine their structural parameters. These spatially resolved light profiles are an important ingredient for the dynamical models of our UCDs. 
A cutout of the \textit{HST} images and the position of the UCDs within CenA is illustrated in Figure \ref{fig:overview}

The point-spread function (PSF) was generated empirically using isolated point sources in \textit{HST} images taken with similar dither patterns. These stars were combined into a single image and a spatially varying PSF was determined using the fortran version of DAOPHOT.

\section{Analysis}
\subsection{Kinematics measurements}

\label{sec:kinmethod}
For our dynamical analysis, we use the strong near-infrared CO band absorption lines from 2.29-2.39\, $\mu$m, which are located in the $K$-band that we observed with SINFONI. To fit stellar templates to the absorption lines, we use the penalized pixel-fitting (pPXF) code (\citealt{Cappellari2004, Cappellari2017}). It allows one to fit a set of model templates to the data and derives the best-fit radial velocity and velocity dispersion of the observed spectrum. For our stellar model spectra, we use the library of high resolution stellar templates of cool stars in the $K$-band from \cite{Wallace1996}. The high-resolution model spectra are convolved with the SINFONI line-spread function, to bring them to the same spectral resolution as our UCD observations.

For UCD\,330, there was sufficient S/N to create 2D kinematic maps using Voroni binning (\citealt{Cappellari2003}). We require that each bin has a minimum $S/N>30$. Outside of $r>0.3\arcsec$ we created bins that span $90^{\circ}$ intervals to maximize the S/N. However, for UCD320 with significantly lower S/N, we needed to restrict our analysis to (1D) radial binning. 

We do not fit the $h3$ (skewness) and $h4$ (kurtosis) parameters, as the spectra do not have the necessary S/N to draw reliable conclusions about the shape of our absorption lines. Before carrying out the kinematic fits we co-add several spaxels into bins to improve the S/N. For the radial dispersion profiles of both UCDs, we add up all pixels in radial bins.

The uncertainties of our kinematic measurements were determined by adding random Gaussian noise to our spectra. The random noise level is based on the residual from the best-fit model. We ran such Monte Carlo simulations 25 times for each spectrum and refitted the kinematics. The standard deviation of the kinematic values from the 25 trials is then adopted as the $1\,\sigma$ kinematic error. 

We perform a barycentric velocity correction for all measured radial velocities. We use the barycentric correction at the date of observation for each individual exposure, and then averaged all corrections. For UCD\,330 the average correction is $v_{\rm bary}=-19.1\,\rm km\, s^{-1}$ and for UCD\,320 it is $v_{\rm bary}=-21.9\,\rm km\, s^{-1}$.

\subsection{Surface brightness and mass profiles}

Every JAM model requires a model for the distribution of the stellar mass within the UCD. We can derive the surface brightness profile of our UCDs from the available \textit{HST} images in the F606W filter. We use the two-dimensional surface brightness code GALFIT (\citealt{Peng2002}) to fit a double S\'ersic light profile. We fitted both UCDs using a $10\arcsec \times 10 \arcsec$ cutout of the \textit{HST} F606W imaging (Fig. \ref{fig:overview}) with an $80\times80$\,pixel PSF convolution box. The best-fit model parameters of the S\'ersic profiles are listed in Table \ref{tab:galfit}. We first run a single S\'ersic profile to measure the best fit center of the UCD. Then we refit a double S\'ersic profile, assuming the same center for both S\'ersic components and keeping it fixed. The other fit parameters, including the magnitude, effective radius, S\'ersic index, ellipticity, and position angle, were all allowed to vary for both UCDs. We also allow GALFIT to account for a background gradient to take into account the varying background light from Cen\,A. For UCD\,320 the single S\'ersic fit was the best fit model, whereas for UCD\,330 the double S\'ersic fits had a lower reduced $\chi^{2}$ value compared to the single component model.

For UCD\,330 we find a best fit inner component with $r_{\rm inner}=0.13\arcsec=2.4$\,pc and a S\'ersic index of $n=1.7$ and $r_{\rm outer}=0.54\arcsec=9.97$\,pc and $n=4.73$ for the outer component. With axis ratios of 0.84 and 0.80 respectively, both components are similar in ellipticity. The combined effective radius of these two components is $r_{\rm eff}=0.2\arcsec=3.69$\,pc, which is larger than the literature value (Table \ref{tab:UCDs}). The total extinction corrected F606W magnitude is $m_{\rm F606}=16.66$ which translates to $m_{\rm V}=16.88$. Thus the absolute magnitude is $M_{\rm V}=-11.03$. 

In \citet{Rejkuba2007} they find $M_{\rm V}=-11.66$ after applying an 0.64\,mag internal extinction correction for dust in Cen\,A, in addition to their external 0.34mag foreground extinction correction. If we only correct for the foreground extinction, the magnitude is $M_{\rm V}=-11.02$, which is consistent with our value. This is the only object for which \citet{Rejkuba2007} applied this additional correction based on the presence of strong NaD lines. However the lines themselves are too noisy to measure the internal extinction directly and thus their internal extinction value is an estimate. In addition the extinction corrected (V-I) colour of UCD\,330 is 0.78 in \citet{Rejkuba2007}, yet for a 12.6\,Gyr old stellar population with the clusters metallicity of $[Fe/H]=-0.4$, the Padova models predict a $(V-I)=1.15$ \citep{Padova2000}; this matches much better when correcting only for the foreground extinction, which yields a $V-I=1.11$ for UCD330. This suggests that their large internal extinction correction is overestimated, and thus we do not apply it.

For UCD\,320 we find the best-fit profile to be a single S\'ersic with $r_{\rm eff}=0.28\arcsec=5.17$\,pc and a S\'ersic index of $n=3.46$ and an axis ratio of 0.65. Thus UCD\,320 is significantly elliptical and smaller than the previous effective radius of 6.83\,pc (Table \ref{tab:UCDs}). It's extinction corrected F606W magnitude is $m_{\rm F606}=17.30$ which translates to $m_{\rm V}=17.52$ and thus the absolute magnitude is $M_{\rm V}=-10.39$ which is the exact value also found in the literature that is also corrected for foreground Milky Way extinction (see Table \ref{tab:UCDs}).

\begin{figure*}
   \centering
   \includegraphics[width=0.9\hsize]{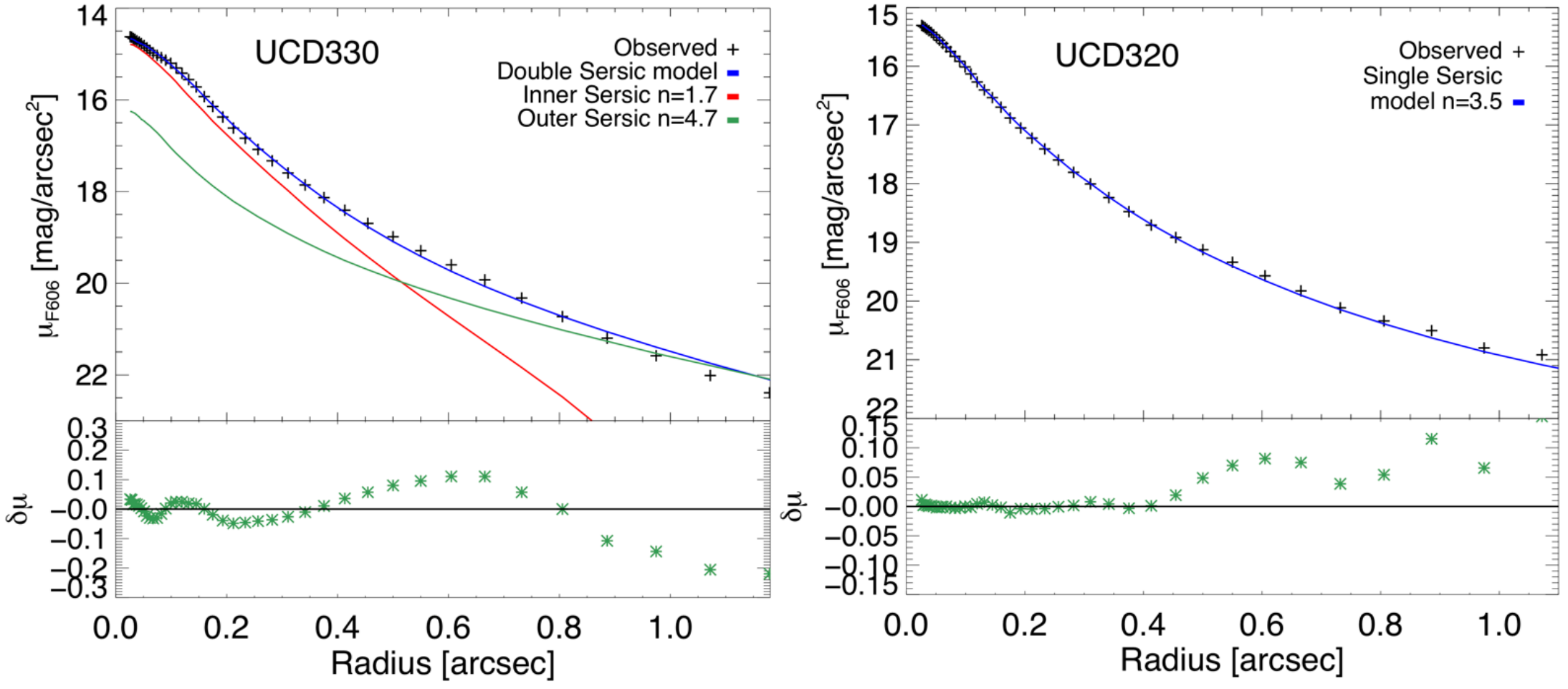}
      \caption{The observed surface brightness profiles of UCD\,330 (left) and UCD\,320 (right) shown as black plus signs. The best fit double S\'ersic model for UCD\,330 is shown as blue line in the left panel and the single S\'ersic model for UCD\,320 in blue un the right panel. For UCD\,330 the inner and outer S\'ersic components are shown individually in red and green respectively. These are the models convolved with the PSF from the \textit{HST} images. The individual S\'ersic parameters can be found in Table \ref{tab:galfit}.}
         \label{fig:surf}
   \end{figure*}

The surface brightness profiles of both UCDs and their best fit S\'ersic models derived with GALFIT are shown in Figure \ref{fig:surf}. The black plus signs are the measured values and the blue line is the best fit model that was convolved with the PSF.

\begin{deluxetable*}{cccccccc}
\tablecolumns{7}
\tablecaption{GALFIT Results. The apparent magnitude is corrected for $A_{\rm V}=0.31$\,mag of extinction. \label{tab:galfit}}
\tablehead{ \colhead{Name} & \colhead{Mag (F606W)} & \colhead{$R_{\rm eff}$ (arcsec)} &  \colhead{$R_{\rm eff}$ (pc)} & \colhead{Sersic Index n} & \colhead{Axis Ratio} & \colhead{P.A.} & \colhead{Reduced $\chi^{2}$}}

\startdata
UCD\,330 Inner S\'ersic & 17.17 & 0.13 & 2.17 &1.70 & 0.84 & $-$48.84 & 4.01 \\
UCD\,330 Outer S\'ersic & 17.72 & 0.54 & 8.97 & 4.73 & 0.80 & $-$48.69 & 4.01 \\
UCD\,330 Single S\'ersic & 16.80 & 0.17 & 3.11 & 1.92 & 0.81 & $-$32.66 & 6.92 \\
UCD\,320  S\'ersic & 17.30 & 0.28 & 4.67 & 3.46 & 0.65 & $-$79.82 & 10.36 \\
\enddata

\end{deluxetable*}

We use the Multi-Gaussian Expansion (MGE) code (\citealt{Cappellari2002}) to parametrize the UCD surface brightness profiles using several two dimensional Gaussian models. The final two dimensional surface brightness model of the UCDs can then be analytically deprojected into a three-dimensional model. The MGE surface brightness profiles (in units of $L_{\odot}/\rm pc^{2}$) are given in Tables \ref{tab:MGE330} and \ref{tab:MGE320} respectively. Assuming a constant mass-to-light ratio for the stellar population means that the surface brightness profile directly translates into the stellar mass profile.

\subsection{Jeans Anisotropic Models}
\label{sec:JAMmodels}

We model our UCDs using the Jeans Anisotropic Models (JAM) code (\citealt{Cappellari2008}) which predict the kinematics of a axisymmetric stellar system based on a supplied luminosity profile. This is compared to kinematic data to constrain the free parameters such as the M/L, BH mass and orbital anisotropy. This code provides both the radial velocity and velocity dispersion parameters as model outputs. In a cylindrical coordinate system where the $z$-axis is aligned with the object's symmetry axis, the anisotropy is defined as $\beta_{\rm z}=1-(\frac{\sigma_{\rm z}}{\sigma_{\rm r}})$.

In addition to the stellar mass profile, the JAM code adds a Gaussian mass profile to model the presence of a central black hole. The JAM models predict the $_{\rm rms}=\sqrt{v_{\rm rad}^{2}+\sigma^{2}}$ profile, and thus cannot be compared to the full LOS velocity distribution as more sophisticated orbit-based models, such as Schwarzschild models, can. Similar to the gravitational effect of a black hole, radial anisotropy raises the dispersion near the center. Thus the black hole mass and anisotropy are intrinsically degenerate with each other.

We can explore the effects of the degeneracy by using grids with a range of anisotropies and BH masses.

For UCD\,330 and UCD\,320, we run a grid of JAM models with four free parameters: black-hole mass $M_{\bullet}$, anisotropy $\beta_{\rm z}$, mass-to-light ratio $M/L_{\rm F606W}$ and the inclination. We use the following grid:
\begin{itemize}
\item 10 values for the BH mass, including a zero mass BH and 9 BH masses ranging from $log(M_{\bullet}/M_{\odot})=4-6.66$ in increments of $log(M_{\bullet}/M_{\odot})=0.33$.
\item 30 values for $M/L_{\rm F606W}$ ranging from 0.55 to 4.9 in steps of 0.15
\item 10 anisotropies $\beta_{\rm z}$ ranging from $-$1.0 to 0.8 in increments of 0.2
\item 6 inclinations ranging from $40^{\circ}$ to $90^{\circ}$ in increments of 10 degrees
\end{itemize} 
In the rest of the paper we do not report the $M/L_{\rm F606W}$ values directly, but rather the quantity $\Psi=\frac{M/L_{\rm F606W}}{M/L_{\rm pop}}$, which is the dynamical mass-to-light ratio normalized with the predicted M/L ratio from stellar population models.

The predicted $M/L_{\rm pop}$ ratios are calculated similar to \cite{Mieske2013}, using the average of the \cite{Maraston2005} and the \cite{BC2003} stellar population models for an object of 13\,Gyr using the mean of a Kroupa and Chabrier IMF. We assume [Fe/H]$=-0.36\pm0.14$ for UCD\,330 and [Fe/H]$=-0.85\pm0.14$ for UCD\,320 taken from \citet{Beasley2008}. Using these metallicities the predicted $M/L_{\rm pop}$ value are $M/L_{\rm pop\,V}=3.30$ and $M/L_{\rm pop\,V}=2.64$ for UCD\,330 and UCD\,320 respectively. We translate this to the F606W band predictions by adopting a V-F606W=0.219\,mag color difference between the $V$ band and the \textit{HST} F606W filter and a 0.1\,mag color difference of the Sun. The F606W predictions are then $M/L_{\rm pop\,F606W}=2.95\pm0.22$ for UCD\,330 and $M/L_{\rm pop\,F606W}=2.37\pm 0.13$ for UCD\,320. The uncertainties on the $M/L_{\rm pop}$ predictions are derived from propagating the 0.14\,dex error of the metallicity measurement (\citealt{Beasley2008}) into the stellar population prediction. 

We used the same grid of JAM models on both UCDs. For each UCD we derive likelihood maps that show the degeneracy between two of the fit parameters each. For this we marginalize the grid over the two parameters that are not plotted. We use the reduced $\chi^{2}$ value of each model to calculate the likelihood for each point in the grid where we evaluated a model. We then use these likelihood values and plot the contours of the 1, 2, and 3\,$\sigma$ levels. The modeling results and likelihood maps are shown in Section \ref{sec:dyn_models}.

 
\section{Kinematic Results} 
\label{sec:results}

\subsection{Integrated velocity dispersions}
First, we co-add the spectra within a circular aperture of $0.4\arcsec$ for UCD\,330 and $0.3\arcsec$ for UCD\,320 to get an integrated spectrum.  Using the pPXF code we fit stellar templates to the observed spectra (see Sec. \ref{sec:kinmethod}) to them to derive their velocity dispersion. 

\begin{figure}
   \centering
   \includegraphics[width=\hsize]{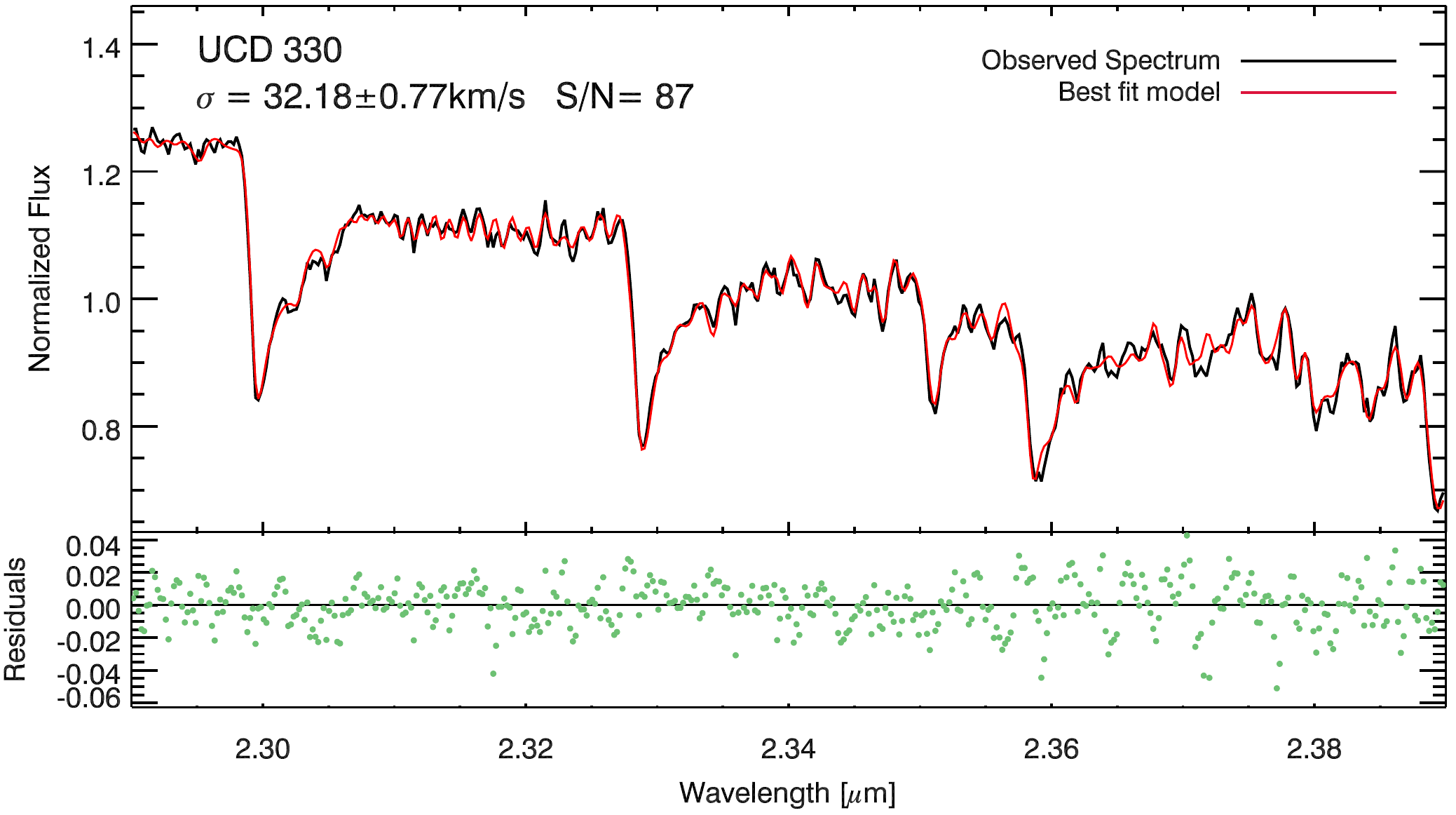}
      \caption{The near infrared spectrum of UCD330 centered on the CO-bandhead absorption features between $2.29\,\mu m$ and $2.39\,\mu m$. The spectrum was integrated out to $0.4\,\arcsec$ in radial distance. The observed spectra is shown in black and the best fit model is plotted in red and the residuals are shown in the panel below in green color. }
         \label{fig:integrated_330}
   \end{figure}

For UCD\,330 the mean signal-to-noise ratio of 87 enables an accurate determination of the integrated velocity dispersion. We find a dispersion of $\sigma_{\rm v}=32.18\pm 0.77\,\rm km\,s^{-1}$ (see Figure \ref{fig:integrated_330}). In comparison the integrated dispersion value of $\sigma_{\rm v}=30.9 \pm 1.5\,\rm km\,s^{-1}$ \cite{Rejkuba2007} was determined from high-resolution UVES data in a 1\,$\arcsec$ aperture. To correct for the larger aperture, we use the JAM models with a pixel size set to 1\,$\arcsec$ to predict the integrated dispersion of UCD\,330 for similar aperture. We predict a dispersion of $\sigma_{\rm v}=30.79\,\rm km\,s^{-1}$ in the larger aperture, which is fully consistent within the errorbars with  \cite{Rejkuba2007} values.

\begin{figure}
   \centering
   \includegraphics[width=\hsize]{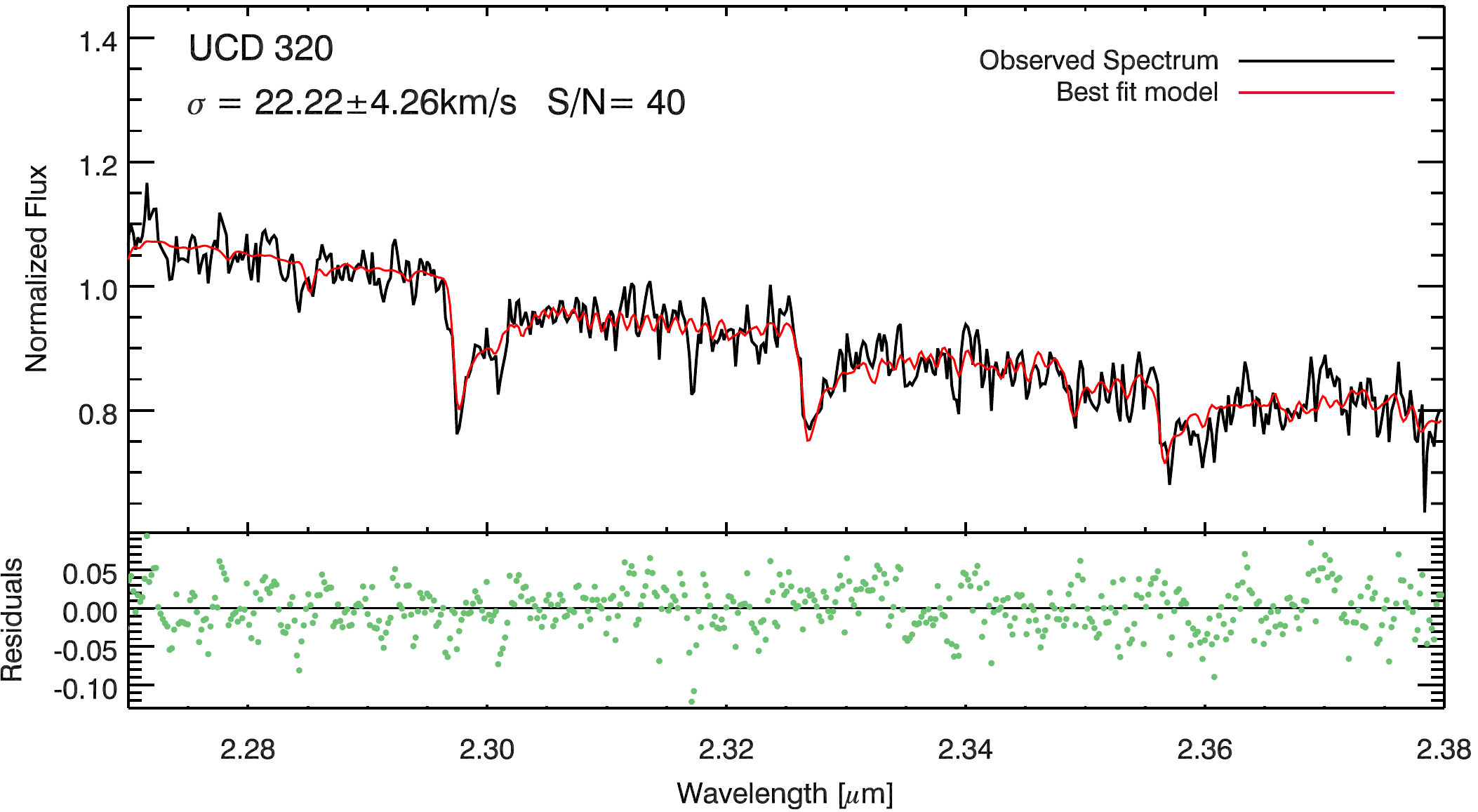}
      \caption{The integrated $K_s$-band spectrum of UCD320 with the CO-bandhead absorption features plotted between $2.27\,\mu m$ and $2.38\,\mu m$. The spectrum was integrated out to $0.3\,\arcsec$ in radial distance, to optimize the S/N that decreases when co-adding more distant spaxels that are noisier due to the decreasing flux of the object. The observed spectrum is plotted in black and the best-fit model is shown in red and the residuals are plotted in the panel below. }
         \label{fig:integrated_320}
   \end{figure}

For UCD\,320 we did a similar analysis, but included the bluer parts of the spectra down to $\lambda = 2.20\,\mu m$ in our fits, to improve the S/N. With this we are able to reach a median S/N of 40 per pixel when integrating out to $0.3\,\arcsec$, which is plotted in Fig.\ref{fig:integrated_320}. Co-adding spectra from spaxels at larger distances does not add to the S/N but rather decreases it. We find an integrated dispersion of $\sigma_{\rm v}=22.22\pm 4.26\,\rm km\,s^{-1}$. Due to the lower S/N for this UCD the measurement has a higher uncertainty.
We also predict the integrated dispersion using a 1\,$\arcsec$ aperture for UCD\,320. We predict a dispersion of $\sigma_{\rm v}=19.70\,\rm km\,s^{-1}$, which is fully consistent with the $\sigma_{\rm v}=20.9\pm 1.6\,\rm km\,s^{-1}$ value derived by \cite{Rejkuba2007} and the $\sigma_{\rm v}=20.0\pm 1.4\,\rm km\,s^{-1}$ measured in \cite{Taylor2010}. The fact that our integrated dispersion is consistent with their value indicates that we can reliably measure velocity dispersions of $\sim20\,\rm km\,s^{-1}$ close to the SINFONI resolution limit.

In \cite{Taylor2010} the high resolution UVES data for UCD\,330 from \cite{Rejkuba2007} were reanalyzed and they found a dispersion of $\sigma_{\rm v}=41.5\pm 3.7\,\rm km\,s^{-1}$ which is a $\sim2.5\,\sigma$ outlier from the integrated dispersions found in \cite{Rejkuba2007} and in this work. A new measurement of $\sigma_{\rm v}=29.2\pm 3.0\,\rm km\,s^{-1}$ of UCD\,330 dispersion from \citet{Hernandez2018} is also consistent with our measurement.

As our dispersion and the original measurement from \cite{Rejkuba2007}, and the independent one of \citet{Hernandez2018} agree with each other, it is likely that the \cite{Taylor2010} value is the outlier. We note that this high dispersion resulted in \citealt{Taylor2010} measuring a dynamical M/L more than twice as high than what was expected. We will revisit this after deriving our best fit M/Ls below.

\subsection{Two dimensional resolved kinematic map of UCD\,330}

\begin{figure*}
   \centering
   \includegraphics[width=\hsize]{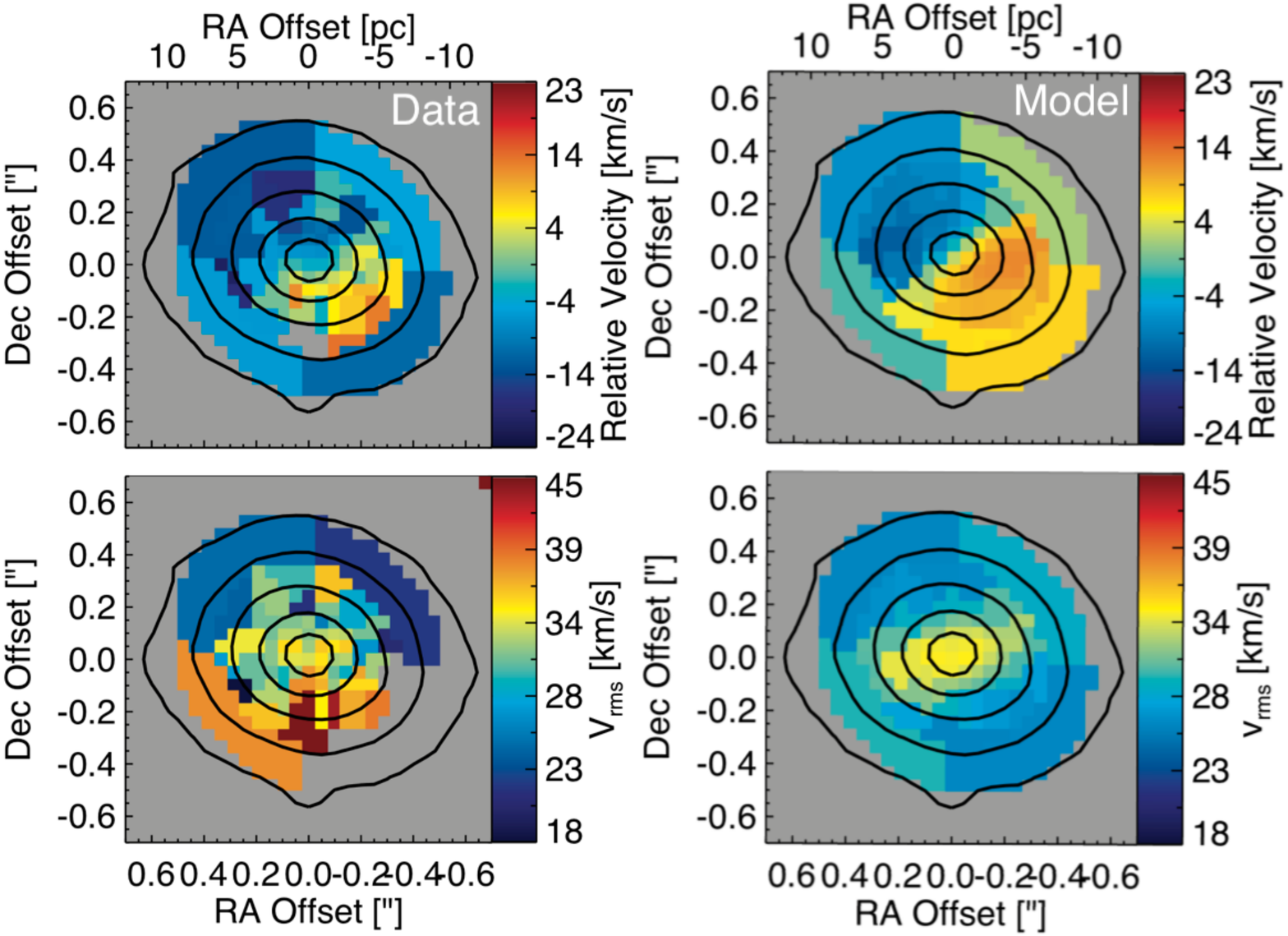}
      \caption{The two dimensional kinematic maps (left) and models (right) of UCD\,330. The top panels show the radial velocities and the bottom panels show the second moment of the LOSVD $v_{\rm rms}=\sqrt{\sigma^{2}+v_{\rm rad}^{2}}$ that includes the velocity dispersion and the radial velocity. The black isophotes show the contours of the stellar light from the $K$-band image. For the central area the signal to noise is high enough that the dynamics were measured for single pixels whereas in the outskirts many pixels were binned together. The typical uncertainties for the $v_{\rm rms}$ are $2\,\rm km\,s^{-1}$ for the central pixels and $6\,\rm km\,s^{-1}$ in the outskirts. Greyed out bins in the data panels are either bins with a $S/N<5$ or the uncertainties on the dispersion and radial velocity are above $15\,\rm km\,s^{-1}$ . }
         \label{fig:JAM_2Dmap}
   \end{figure*}

The high quality of the UCD\,330 data permits us to measure a resolved two dimensional kinematic map. The results of the kinematic measurements are shown in the two left panels of Figure \ref{fig:JAM_2Dmap}. The map of the radial velocity is shown in the top left panel and the map of the second order momenta $v_{\rm rms}=\sqrt{v_{\rm rad}^{2}+\sigma^{2}}$ is shown in the bottom left panel. The best fit JAM model is shown in the two panels on the right and we will discuss its results in Section \ref{sec:dyn_models}. The typical uncertainties on the $v_{\rm rms}$ are $2\,\rm km\,s^{-1}$ for the central individual pixel bins, and $\sim 6\,\rm km\,s^{-1}$ in the outer larger bins.

The observed velocity map is normalized to the systemic velocity of UCD\,330 of $v_{\rm sys}=743\,\rm km\,s^{-1}$. The amplitude of the observed rotation is $\sim12\,\rm km\,s^{-1}$ with the rotation axis aligning with the semi-minor axis of the UCD. 
The fraction of rotational versus dispersion support in this UCD is $v_{\rm rot}/\sigma=0.37$, when comparing it to the global velocity dispersion. This indicaties a significant contribution from rotation. With an average axis ratio of 0.82 for UCD\,330 we would expect $v_{\rm rot}/\sigma=0.4$ (\citealt{Binney1978}) from a self gravitating system that is flattened by its rotational support, which is consistent with what we observe.

The observed $v_{\rm rms}$ map in the bottom left panel of Fig. \ref{fig:JAM_2Dmap} is more complex. In the top half of the map the dispersion in the outskirts is  $\sim20\,\rm km\,s^{-1}$ and increases towards values of  $\sigma = 34\,\rm km\,s^{-1}$ in the center with observational noise adding some scatter. However the top half of the dispersion map is overall consistent with a radially decreasing velocity dispersion profile.

The velocity dispersion at the bottom of the $v_{\rm rms}$ map appears unusual with several high-dispersion outer bins. An asymmetric dispersion profile is highly unusual as the dispersion is expected to decrease outwards at all angles. It is unclear what could cause such a high-dispersion region. We tested whether it could be a detector issue, by analyzing the individual cubes separately and checking if the results are consistent. We did not find a significant difference between the kinematic results for individual cubes.
As our individual cubes are dithered by significant amounts so that in each individual cube the UCD is at a different detector location, it is unlikely to be a detector effect. A physical explanation could be that the UCD is semi-resolved into stars and internal bright-star variations causing the elevated dispersion, or that an object in projection contaminates the measurements. Another alternative is that the UCD is tidally disturbed and that increased its dispersion. However, in this case it would be strange that the disturbances are confined  confined to only one side of the UCD and are not symmetric.

 \subsection{Radial Dispersion Profiles}
 
\begin{deluxetable}{ccccc}
\tablecaption{Radial velocities and velocity dispersion of UCD330 and 320 and the signal-to-noise of each radial bin \label{tab:Radial}}
\tablehead{\colhead{Radius} & \colhead{Radius} & \colhead{Velocity $v_{r}$} & \colhead{Dispersion $\sigma_{v}$} & \colhead{S/N} \\ 
\colhead{(arcsec)} & \colhead{(pc)} & \colhead{(km/s)} &  \colhead{(km/s)} & \colhead{} } 

\startdata
\centering UCD\,330 & & & & \\ 
\hline
0.025 & 0.42 & 735.8 $\pm$ 0.90 & 33.98 $\pm$ 0.84 & 83.8 \\ 
0.075 & 1.25 & 737.8 $\pm$ 0.74 & 34.02 $\pm$ 0.73 & 104.3  \\
0.125 & 2.08 & 739.8 $\pm$ 0.75 & 34.03 $\pm$ 0.75 & 102.0  \\
0.175 & 2.91 & 741.8 $\pm$ 0.92 & 32.45 $\pm$ 1.01 & 79.9  \\
0.225 & 3.74 & 741.7 $\pm$ 1.46 & 34.10 $\pm$ 1.41 & 56.8   \\
0.275 & 4.57 & 738.8 $\pm$ 1.93 & 33.30 $\pm$ 2.08 & 42.0   \\
0.350 & 5.81 & 736.9 $\pm$ 2.32 & 24.38 $\pm$ 2.92 & 29.2   \\
0.450 & 7.48 & 742.7 $\pm$ 4.96 & 31.86 $\pm$ 5.78 & 17.9   \\
\hline
\centering UCD\,320 & & & & \\ 
\hline
0.025 & 0.42 & 500.8 $\pm$ 5.3  & 29.52 $\pm$ 11.57 &  28.50 \\
0.075 & 1.25 & 498.2  $\pm$  3.3 & 21.05 $\pm$ 5.22 & 36.03 \\
0.125 & 2.08 & 497.9  $\pm$   3.7 & 20.64 $\pm$ 4.46 & 33.29 \\
0.175 & 2.91 & 500.5  $\pm$   3.4 & 21.69 $\pm$  4.37 & 34.46 \\
0.250 & 4.15 & 505.5 $\pm$    5.0 & 22.82 $\pm$ 7.52 & 23.76 \\
\enddata




\end{deluxetable}

\begin{figure}
   \centering
   \includegraphics[width=\hsize]{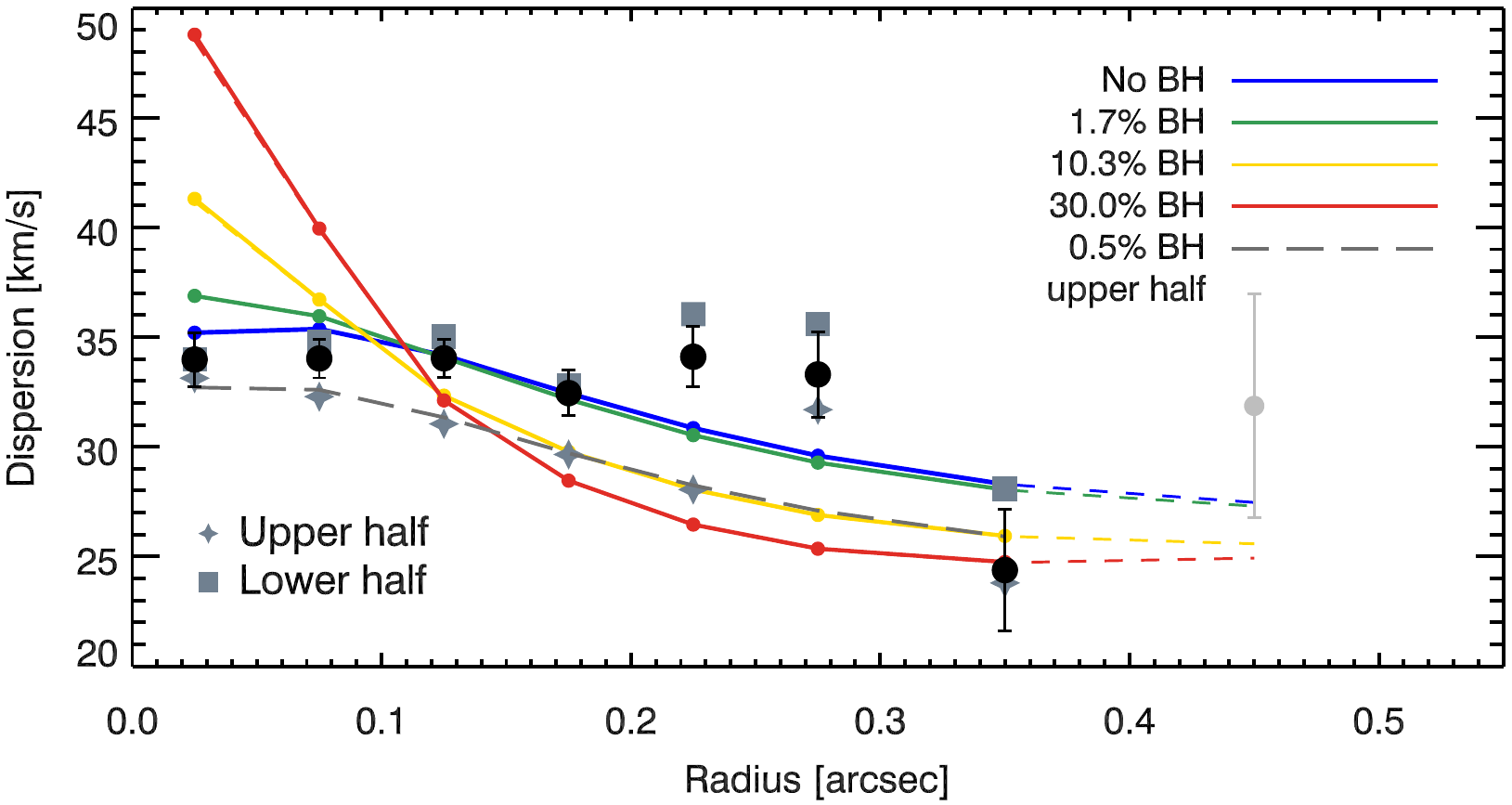}
      \caption{Black points are the measured radial dispersion profile of UCD330. The colored lines are the best fit isotropic models with increasing BH mass, the dashed coloured lines are the models extended to the region where we do not fit them. The zero mass BH model (blue) is the best overall model and the model with 1.7\% BH mass fraction (green) is the $3\,\sigma$ upper limit. The BH fractions of 1.7\%, 10.3\%, 30.0\%  correspond to absolute BH masses of $1.0\times10^{5}, 4.6\times10^{5}, 1.0\times10^{6}\,M_{\odot}$. The light grey datapoint of the outermost bin was not included in the JAM modeling. The dark grey squares and star symbols are the radial dispersion profile of the lower and upper half of the UCD respectively. The dashed grey line is the best-fit upper-half model containing a BH of $2.14\times10^{4}\,M_{\odot}$ corresponding to a BH fraction of 0.5\%.}
         \label{fig:JAM_iso_330}
   \end{figure}

In addition to the two dimensional kinematical map of UCD\,330 we also measure the radial dispersion profile. The radial dispersion values can be found in Table \ref{tab:Radial} and are plotted as black points in Figure \ref{fig:JAM_iso_330}. For UCD\,330 we have high signal-to-noise data and thus were able to use 7 radial bins of $0.05\,\arcsec$ width to measure the kinematics. The S/N for our central bins is between 80-100 and decreases to 20-40 in the outskirts. Our measured velocity dispersion errors are small with $1\,\rm km\,s^{-1}$ in the center and $2\,\rm km\,s^{-1}$ in the outskirts.

For UCD\,330 the resulting radial dispersion profile in Figure \ref{fig:JAM_iso_330} is very flat.

From the two dimensional map in the previous section we have indications that the increased dispersion values in the UCD come from the increased region at the bottom of the UCD. 

To investigate the differences between the 'upper' and 'lower' part of the UCDs, we split its velocity dispersion profile with a horizontal line crossing the center and remeasure the radial dispersions separately. These are plotted in Figure \ref{fig:JAM_iso_330} as grey squares and stars respectively. Comparing the radial dispersion profiles of both halves has the advantage that we get a higher signal-to-noise per bin compared to the two dimensional map to test if the high dispersion area in the 2D map are significant.

It is apparent from Figure \ref{fig:JAM_2Dmap} that the flat and even rising dispersion is mainly caused by the lower half of UCD\,330 (grey squares) which is much higher in the outskirts than the dispersion measurements for the upper half (grey stars) that are gradually decreasing towards the outskirts, with one outlier from this gradual decrease at the $0.275\, \arcsec$ bin.

Considering the statistical significance of the differences between 'upper' and 'lower' velocity dispersion values we find that the lower half measurements at $r=0.075,0.125,0.175,0.225\arcsec$ are discrepant with respect the measurements of the upper UCD half (grey star symbols) at 2.5, 3.2, 1.9 and 3.8\,$\sigma$ significance. Taken together, it is clear that this enhancement is significant. The flatness of the radial dispersion profile in the outskirts (see Fig. \ref{fig:JAM_iso_330}) is caused  mostly by the contribution of the high dispersion region in the lower half.

   \begin{figure}
   \centering
   \includegraphics[width=\hsize]{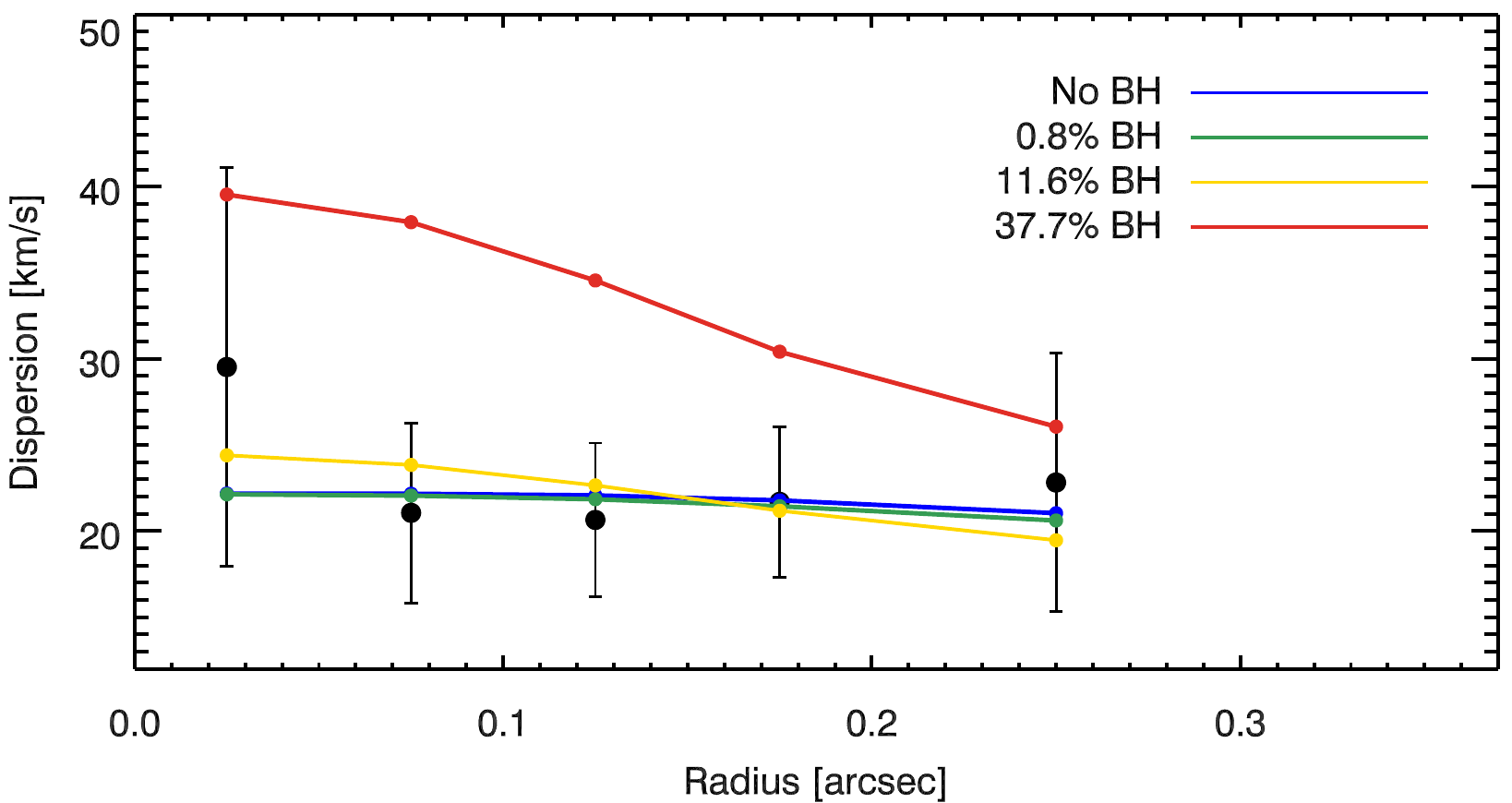}
      \caption{Black points are the measured radial dispersion profile of UCD\,320. The colored lines are models models with increasing BH mass. The 0.8\% BH model (green) is the best overall model and the 37.7\% model (red) is the $3\,\sigma$ upper limit. The BH fractions of 0.8\%, 11.6\%  and 37.7\% correspond to absolute BH masses of $2.15\times10^{4}, 2.15\times10^{5}$ and $1\times10^{6}\,M_{\odot}$. }
         \label{fig:JAM_iso_320}
   \end{figure}

The radial dispersion profile of UCD\,320 is much less well constrained due to the poorer data quality. We have 5 radial bins for which we are able to measure the radial velocity and velocity dispersion (see Table \ref{tab:Radial}), which is plotted in Figure \ref{fig:JAM_iso_320} as black points. Those bins reach a S/N ratio between 24-36. The dispersion profile is flat for the outer 4 bins with values around $\sigma_{\rm v}=20-22\,\rm km\,s^{-1}$, which is consistent with what we derived for the integrated dispersion. Only the central bin shows an increase in dispersion to almost $30\,\rm km\,s^{-1}$ but with the large uncertainty of $11\,\rm km\,s^{-1}$, this increase is not statistically significant.
 
We measure the rotation of UCD\,320 by dividing the data at varying position angles and measure the radial velocity for both sides. This is done in intervals of $10^{\circ}$ in PA. The resulting amplitude A, of the rotation curve is $A=5.25\,\rm km\,s^{-1}$. Taking into account that the rotational velocity varies with the azimuthal angle we calculate the true rotation velocity using: $v_{\rm rot}=A/\frac{\pi}{4}=6.68\,\rm km\,s^{-1}$. Therefore we derive the rotational versus dispersion support in this UCD as $v_{\rm rot}/\sigma=0.3$. For its axis ratio of 0.65 we expect a $v_{\rm rot}/\sigma=0.6$ (\citealt{Binney1978}) from a self gravitating system that is flattened by its rotational support, which is significantly higher than the measured value, therefore some anisotropy is implied.

 \section{Dynamical Models}
\label{sec:dyn_models}
 
\subsection{Two dimensional JAM models for UCD\,330}  
We run a large grid of two dimensional JAM models (Section \ref{sec:JAMmodels}) that allow for varying black hole masses, anisotropy, mass-to-light ratios and inclination angles. The JAM model predictions are used to fit the $v_{\rm rms}$ data shown in Fig.~\ref{fig:JAM_2Dmap}.

No black hole is detected in UCD\,330.  The best-fit model has a black hole mass of 0 with a $3\,\sigma$ upper limit of $1.0\times10^{5}M_{\odot}$ which equals 1.7\% of the best-fit total mass. We quote 3$\sigma$ errors on all model quantities. The best fit $M/L$ is 2.65 in the F606W band, which translates to an $M/L_{V}$ of 2.97. Our best-fit model is isotropic with $\beta_{\rm z}=0.0^{+0.2}_{-0.4}$. The best fit model is shown in Fig. \ref{fig:JAM_2Dmap} along with the data. The central rise in the 2D model up to $35\,\rm km\,s^{-1}$ is similar to what is observed in the UCD in the upper half. The increased dispersion in the lower of half of UCD\,330 is not reproduced by our model whose velocity dispersion decreases with lager radius. In the upper half of the UCD we reproduce the observed velocity dispersion levels in the outskirts of $\sim24\,\rm km\,s^{-1}$  In addition to the $v_{\rm rms}$ data to which the JAM models were fit, the top panels of Fig.~\ref{fig:JAM_2Dmap} also show the velocity field data and predictions from this best fit model assuming the rotation parameter of $\kappa = 1$ \citep{Cappellari2008}. The observed rotational amplitude of $\sim12\,\rm km\,s^{-1}$, which is aligned with the semi-minor axis of the UCD, is well reproduced in the best-fit JAM model.

We compare the predicted $M/L_{F606W pop} = 2.95 \pm 0.22$ to the measured dynamical M/L (Section 3.3) and find a $\Psi=\frac{M/L_{\rm dyn}}{M/L_{\rm pop}}=0.90^{+0.27}_{-0.55}$, meaning that the dynamical mass is similar to the prediction from stellar population models within the errorbars. This is significantly lower than the $\Psi_{330}=\frac{M/L_{\rm dyn}}{M/L_{\rm pop}}=2.28$ that was determined previously in \cite{Taylor2010}.

We derive a total UCD mass of $M_{\rm tot}=6.1\pm0.23 \times10^{6}M_{\odot}$.  This is significantly smaller than the dynamical mass $1.4\times10^{7}M_{\odot}$ estimated with a single integrated dispersion value by \citet{Taylor2010}. This difference is caused by our $10\,\rm km\,s^{-1}$ lower dispersion value, which agrees with \cite{Rejkuba2007}. The $3\,\sigma$ BH limit of $M=1\times10^{5}M_{\odot}$, making up 1.7\% of the UCD mass, excludes a SMBH of a high mass fraction at high significance.

\begin{figure*}
   \centering
   \includegraphics[width=\hsize]{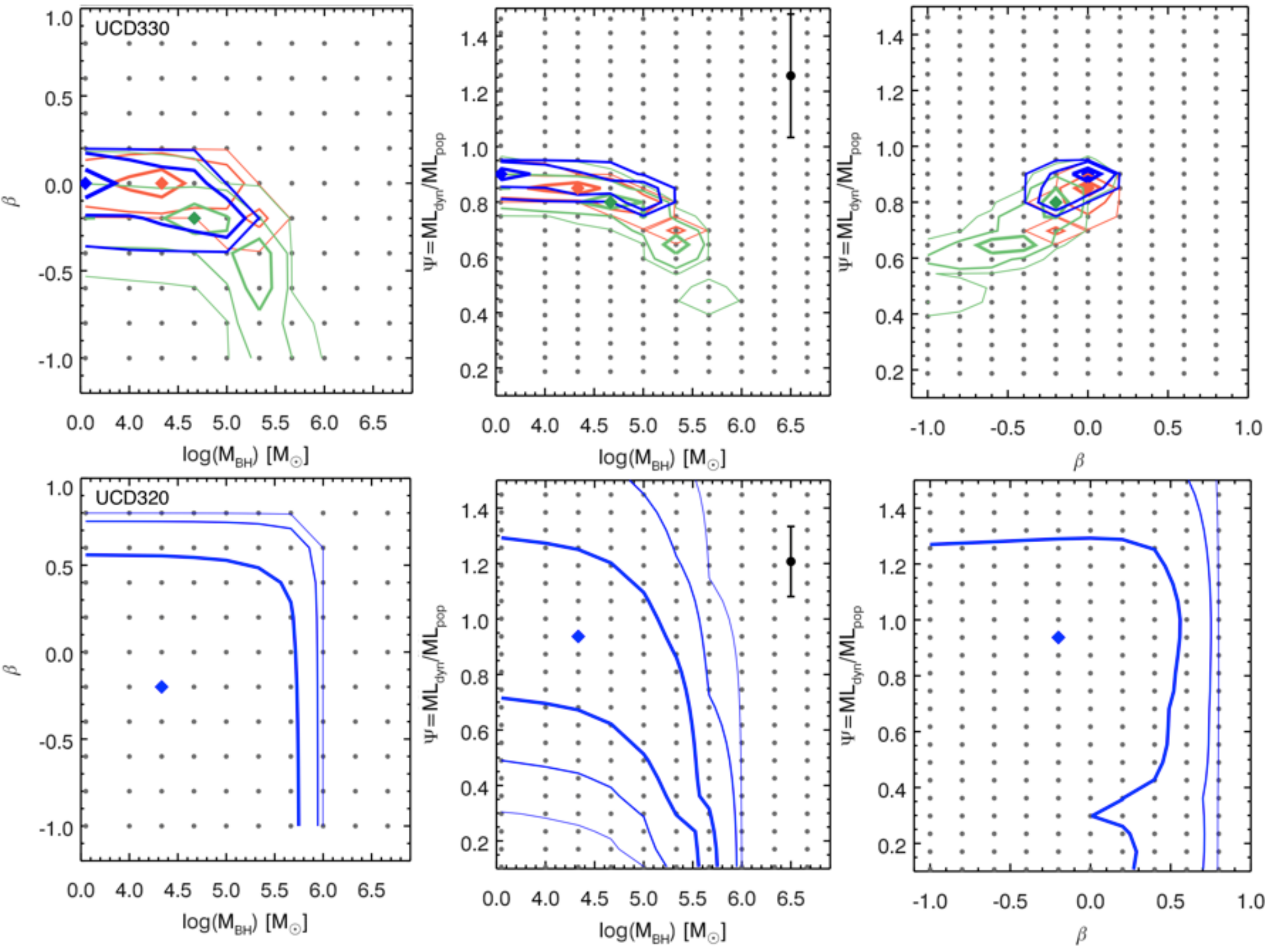}
      \caption{The two dimensional likelihood contours show the dependence of anisotropy $\beta_{\rm z}$, black hole mass $M_{\bullet}$ and the normalized M/L ratio $\Psi$. The top row shows the panels for UCD\,330 and the bottom row for UCD\,320. In blue the likelihood contours for the 2D model are shown, where the 1, 2 and $3\,\sigma$ confidence levels have decreasing line thickness. For UCD 320 the blue contours show the results for model fits to the radial profile. The set of green contours shows the likelihood map for the radial dispersion profile for UCD\,330 and the orange contours when just fitting the upper-half of the 2D dispersion map. The grey points mark each JAM model of the grid and the blue and green diamonds are the best-fit model. The error bar in the middle panel of both UCDs indicates the systematic error in $M_{\rm pop}$ and thus $\Psi$ due to the uncertainty of the metallicity.}
         \label{fig:2dmap}
   \end{figure*}

The two dimensional likelihood distribution comparing 2 parameters each, are shown as the blue contours in the top panel of Figure \ref{fig:2dmap}. Each plot is marginalized over the other parameters that are not plotted. The parameters show minimal covariance, but at the highest allowed BH masses the best fit models have $\sim$10\% lower $M/L$s ($\Psi \sim 0.8$) and prefer $\beta_{\rm z}$ values of roughly $-$0.2. Shown as red contours in Fig. \ref{fig:2dmap}, are the constraints when fitting only the upper part of the 2D dispersion map. We find a $3\,\sigma$ BH limit of $M_{\bullet}=2.15\times10^{5}M_{\odot}$ equal to a 4.3\% mass fraction BH. The upper limit on the BH mass is somewhat larger, mostly due to a lower M/L in the 2D upper half models, but generally the contours are consistent. Our overall conclusions that there is no massive BH in UCD\,330 does not change whether we use the upper-half or the entire 2d dispersion map, indicating that our models are robust.

In the next section, we consider models fit just to the radial profile; these are shown as green contours in Fig. \ref{fig:JAM_2Dmap}. Those models have wider uncertainties due to smaller number of data points fit. We therefore adopt the best-fit models to the 2D data and the errors from these fits as our final values.

To assess our level of systematic error due to the poorly constrained PSF, we also reran the dynamical models with the upper or lower limit on the size and light fraction of the kinematic PSF. The likelihood contours were essentially unchanged, showing that our results are robust to small variations in the kinematic PSF.

\subsection{UCD\,330 radial JAM models}
In this section, we discuss JAM models fit to the radial profile of UCD\,330 only (black points in Fig. \ref{fig:JAM_iso_330}), as a comparison to the two 2D from the previous section. We run them over the same grid as detailed in section \ref{sec:JAMmodels}. The best contours for the grid of radial models is shown in green in the top row of Fig.\ref{fig:2dmap}. 

From the top left panel it is visible that the best fit radial model for UCD\,330 is tangentially anisotropic with $\beta_{\rm z}=-0.2$ and the best-fit BH mass is not 0 anymore but $M_{\bullet}=4.64\times10^{5}M_{\odot}$. The shift towards more tangential anisotropy is the main difference between the radial and the 2D models, but their likelihood contours still overlap. Using only the radial fits the $3\,\sigma$ upper limit on the BH mass would increase to $M_{\bullet}=4.64\times10^{5}M_{\odot}$ equal to 10.3\% of the total mass (yellow model, Fig. \ref{fig:JAM_iso_330}). However, as the radial model constraints are weaker than those from the 2D model due to the larger number of data points in the 2D models. Therefore we use the results from the 2D models as our final BH constraints for UCD\,330.  The increased tangential anisotropy is likely due to the high dispersion area in the lower sections of the UCD, which are integrated into the radial profile and cause it to be flatter than it would be without that area.

To help visualize our BH mass limits, we compare a set of radial isotropic models with different black hole masses to the radial profile of UCD\,330 in Figure~\ref{fig:JAM_iso_330}. This plot shows that the best-fit 2D case without a BH (blue line) fits the 1D data very well. The only significant outliers to that model are the two data points in the outskirts, and those are likely due to the high-dispersion region in the lower parts of the UCD. Excluding the higher dispersion portions of the UCD (the ``upper half'' data plotted as grey stars in Fig.~\ref{fig:JAM_iso_330}), we find particularly good agreement with the shape of a 0.5\% BH model (dashed grey line) with a lower M/L value, but the BH mass is consistent with zero within 2$\sigma$.

The maximum mass BH plotted here is $1\times10^{6}M_{\odot}$ with a 30\% mass fraction, its steep rise in the center is clearly excluded by the data.

\subsection{Radial dynamical models for UCD\,320}

For UCD\,320, our lower signal-to-noise data means we cannot construct a 2D map, and instead we only consider the radial dispersion profile.  As with UCD\,330, we also find no evidence for a BH, however, our constraints on the allowed BH mass are much weaker than for UCD\,330. We model the UCD\,320 radial dispersion data with two sets of models; first we examine the results from a grid that includes the full range of anisotropy, then we look at a comparison of isotropic models to the radial dispersion.

Results from the full grid of JAM models for UCD\,320 that include anisotropy are shown in the lower panel of Fig.\ref{fig:2dmap}.  We derive a best-fit black hole mass of $M_{\bullet}=2.15\times10^{4}M_{\odot}$ which corresponds to 0.8\% of the total UCD mass. The $3\,\sigma$ upper limit of $1\times10^{6}M_{\odot}$ corresponds to a 37.7\% mass fraction. This BH mass limit is independent of the assumed anisotropy in the models.
Our best-fit $M/L_{\rm F606 dyn}$ is $2.20^{+1.9}_{-1.0}$.  From this, we derive a total UCD mass of $M_{\rm tot}=2.81^{+.2.4}_{-1.3} \times10^{6}M_{\odot}$, and find $\Psi=\frac{M/L_{\rm dyn}}{M/L_{\rm pop}}=0.94^{+0.8}_{-0.5}$ (these are $1\,\sigma$ errorbars, including the systematic uncertainty in $M/L_{\rm pop}$ of 0.13) indicating that the dynamical mass of this model is similar to what is predicted from stellar population models. The best-fit anisotropy is $\beta_{\rm z}=-0.2$ with a $3\,\sigma$ upper limit of $\beta_{\rm z}=0.8$, but as visible in Fig. \ref{fig:2dmap} the anisotropy is not tightly constrained and thus no lower limit for $\beta_{\rm z}$ can be determined.

For the zero BH mass model, we derive a stellar mass of $2.98\times10^{6}M_{\odot}$ of the UCD, which is $\sim50\%$ lower than the $6.3\times10^{6}M_{\odot}$ derived from the virial mass estimate based on the integrated dispersion from \cite{Taylor2010}. Given that our integrated dispersion is consistent with theirs, this discrepancy appears to be due primarily to our mass modeling. Specifically, we find an effective radius of $r_{\rm eff}=5.17$\,pc, smaller than their $r_{\rm eff}=6.8\,\rm pc$. Additional differences include their assumption of a virial factor compared to our more accurate MGE mass modeling. As noted above, with our best-fit $\Psi=0.94$, our mass measurement places UCD\,320 among the UCDs without elevated dynamical masses.

We plot the results of the isotropic radial JAM models for UCD\,320 in Figure \ref{fig:JAM_iso_320} for several BH mass contributions. 
The best-fit BH mass model of 0.8\% ($M_{\bullet}=2.15\times10^{4}M_{\odot}$) is shown in green and the $3\,\sigma$ upper limit on the BH mass of 37.7\% ($1\times10^{6}M_{\odot}$) is shown in red. Although we observe an increase of the dispersion in the very central bin, due to its large errorbar it is not significant. The rest of the dispersion profile is very flat at just above $20\,\rm km\,s^{-1}$.


\section{Discussion}
\label{sec:discussion}

   \begin{deluxetable}{lcc}
\label{tab:summary}
\tablecaption{Summary of the measured values and limits for UCD 330 and UCD320}
\tablehead{\colhead{} & \colhead{UCD\,330} & \colhead{UCD\,320} } 
\startdata
$M_{\rm V}$ & $-11.03$ & $-10.39$  \\
$M_{\rm tot}$ \,\, $[M_{\odot}]$ & $6.10\pm 0.23 \times10^{6}$ & $2.81^{+2.5}_{-1.3}\times10^{6}$ \\ 
$\Psi=M_{\rm dyn}/M_{\rm pop} $ & $0.90^{+0.3}_{-0.6}$ & $0.94^{+0.8}_{-0.5}$  \\
$M/L_{\rm F606}$ & $2.65^{+0.15}_{-0.55}$ & $2.20^{+1.9}_{-1.0}$  \\
$M/L_{\rm V}$ & $2.97^{+0.17}_{-0.50}$ & $2.47^{+2.1}_{-1.0}$  \\
$\beta_{\rm z}$ & $0.0^{+0.2}_{-0.4}$ & $-0.2^{+0.6}$  \\
$3\,\sigma$ upper limit: $M_{\bullet}$ \,\,  $[M_{\odot}]$& $1.0\times10^{5}$ & $1.0\times10^{6}$   \\
$3\,\sigma\,$upper limit $M_{\bullet}/M_{\rm tot}$\,\, [\%] & 1.7 & 37.7  \\
\enddata




\end{deluxetable}

In this work we have tested whether two lower mass UCDs in CenA host a SMBH in their centers (\citealt{Taylor2010, Mieske2013}), which would imply that they are the stripped nuclear star clusters of dwarf galaxies. From our dynamical modeling of adaptive optics kinematic data we find that (1) no BHs are detected, and (2) no elevation in the mass-to-light ratios was found even for models with no BH, contrary to previous measurements  \citep{Taylor2010}. In this section, we discuss these results in context of previous results and the implications for the formation of UCDs.

\subsection{Upper limits on BH masses}
We find that a BH larger than $M_{\bullet}=1.0\times10^{5}M_{\odot}$ is excluded at a $3\,\sigma$ confidence in UCD\,330. This corresponds to 1.7\% mass fraction. The dynamical models for UCD\,320 place a $3\,\sigma$ upper limit on a BH in UCD\,320 at $M_{\bullet}=1.0\times10^{6}M_{\odot}$, which corresponds to a BH mass fraction of 37.7\%. 

We have measurements of BH mass fractions of 13\%, 15\%, and 18\% and in three massive UCDs \citep{Seth2014, Ahn2017}. The 1.7\% $3\,\sigma$ limit on the BH mass fraction in UCD\,330  is significantly below these typical BH mass fractions. Thus it is clearly different from these massive UCDs with high mass fraction BHs. In UCD\,320 our limit on the BH mass fraction is much higher and the 37.7\% indicates that a high mass-fraction BH similar to those that have been previously discovered is still allowed by our models.

There is evidence that UCD\,320 hosts an X-ray source with an ultra-luminous peak flare luminosity of $9^{+4}_{-3}\times10^{39}\,\rm erg\,s^{-1}$ (\citealt{Irwin2016}). While the sustained X-ray luminosity is consistent with a normal X-ray binary, the flaring luminosity could suggest a massive BH as it is brighter than the typical X-ray binary luminosities (\citealt{She2017}). 
Assuming that this flare is caused by an accreting BH, the X-ray flare timing places and upper limit of $2\times10^{6}M_{\odot}$ on the BH mass. While the exact reason for the flare is unknown, their BH upper limit is consistent with ours and thus a massive BH in UCD\,320 is still possible.

     \begin{figure}
   \centering
   \includegraphics[width=\hsize]{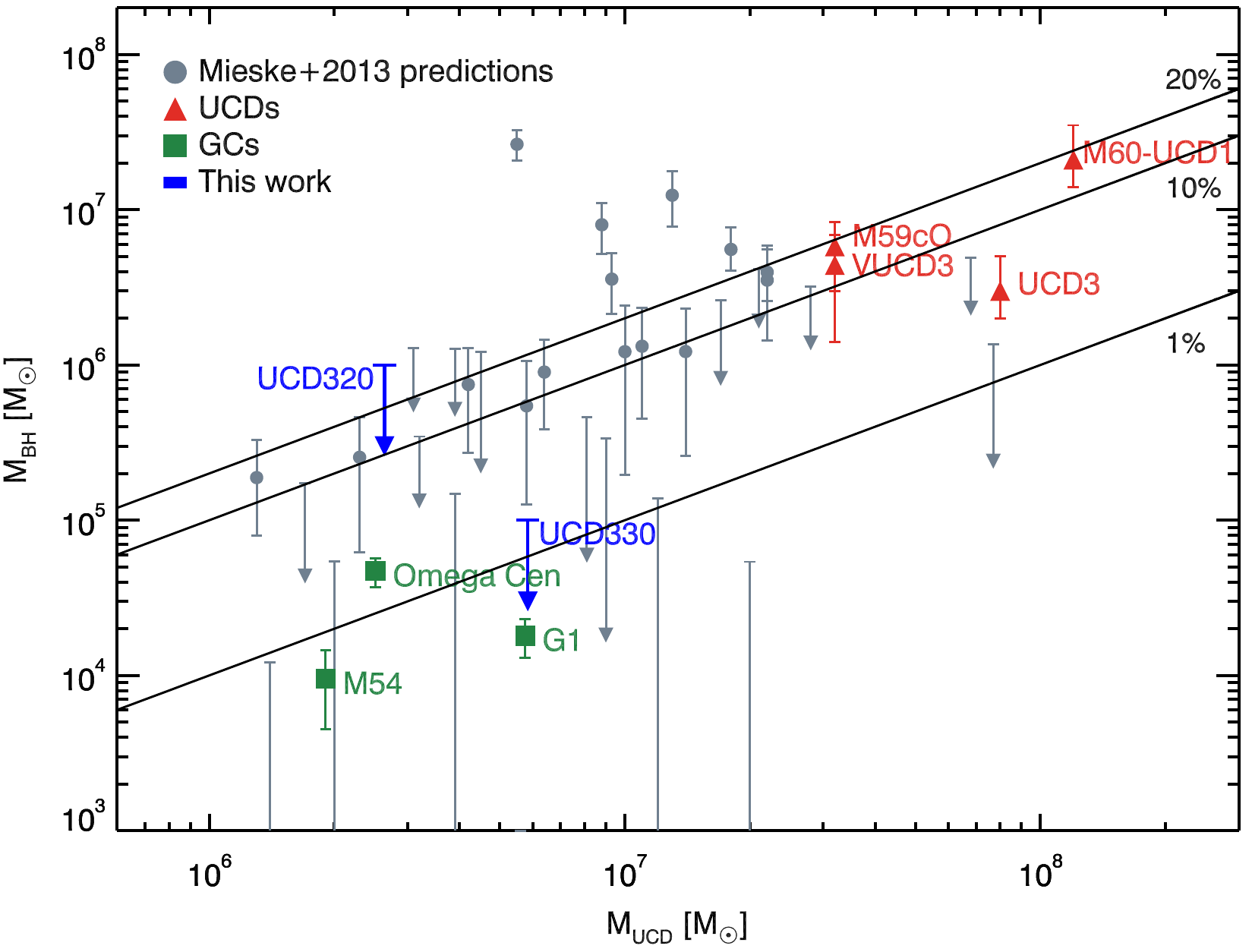}
      \caption{Comparison between UCD and GC masses and their central BHs. Grey points are the predictions and upper limits for UCDs based solely on their integrated dispersions from \cite{Mieske2013}, Green squares are the measured BHs in GCs $\omega$\,Cen, G1 and M\,54 taken from \cite{Lutzgendorf2013}. The 4 known BH masses in UCDs are plotted as red triangles taken from \cite{Seth2014, Ahn2017}, Afanasiev et al. submitted. The three black lines mark constant fractions of 1\%, 10\% and 20\% BH mass fraction respectively.}
         \label{fig:massmbh}
   \end{figure}

In Fig. \ref{fig:massmbh} we compare existing BH mass measurements in GCs and UCDs to the upper limits in both our objects. We also added the predicted BH masses from \cite{Mieske2013} based on assuming that any $M/L$ enhancement observed in a given UCD from integrated dispersion measurements is due to a BH. The black lines are a constant BH mass fraction of 1\%, 10\% and 20\% respectively. Three of the massive UCDs (red) with measured BH masses are within the 10-20\% range. However, the possible intermediate-mass BHs detected in Local Group GCs show mass fractions of $<$2\%. Although these BH masses have been measured in several Local Group GCs (\citealt{Ibata2009, Noyola2010, Lutzgendorf2013, Baumgardt2017}), the BH signal is intrinsically degenerate with significant radial anisotropy and thus some of the detections remain controversial (\citealt{vanderMarel2010, Zocchi2017}).
The upper limit of UCD\,320 is consistent with a such a 10-20\% BH mass fraction, and thus is not constraining, but the in UCD\,330, only a source similar to the local group GCs intermediate-mass BHs could be present. We note that including all the data from the \citet{Mieske2013}, there seems to be a trend of lower mass fraction BHs in lower mass UCDs; our upper limits are fully consistent with this trend.

\subsection{Mass-to-light ratios}

In this work we have shown that contrary to previous findings, UCD\,320 and UCD\,330 have M/L ratios that are not inflated. In both cases, our best-fit models are completely consistent with their stellar populations within the errors, at $\Psi=\frac{M_{\rm dyn}}{M_{\rm pop}}=0.9$.

In this context, the lack of detectable BHs in these systems is not surprising, as in a majority of the more massive UCDs with detected BHs, the best-fit no BH mass models have $\Psi > 1$ \citep{Mieske2013,Seth2014,Ahn2017}. This suggests that finding an inflated $M/L$ does appear to be a strong indicator of the presence of a BH.

At the same time, our results point to the challenges of accurately measuring the dynamical $M/L$s from integrated dispersions. In UCD\,330, the higher velocity dispersion of 41.5\,km/s found by \cite{Taylor2010} yielded an $M/L_V$ of 6.3$^{+1.6}_{-1.7}$, giving a $\Psi=2.3$. Remodeling of this cluster using the \citet{Taylor2010} dispersion by \citet{Mieske2013} yielded a somewhat lower, but still significantly inflated $\Psi=1.7$.  Our much lower $\Psi=0.9^{+0.3}_{-0.6}$ (and $M/L_V =2.97^{+0.2}_{-0.5}$) value results primarily from our lower dispersion, but may also be due in part to our two-component mass model.  As discussed above, our lower dispersion is consistent with the previously published dispersion of \citet{Rejkuba2007} based on analysis of the same data as presented in \citet{Taylor2010}.

In UCD\,320, we also find a much lower value than previous measurements, but for different reasons. Taylor measured an $M/L_{\rm V} =  5.3^{+0.8}_{-1.0}$; because of the lower metallicity of this system relative to UCD\,330, this resulted in an even higher estimate of $\Psi = 2.5$; while the \citet{Mieske2013} analysis found $\Psi=1.6$.  In this case, our lower $\Psi$ value appears to come in part from from our smaller derived effective radius of 5.2\,pc, while \citet{Taylor2010} use a significantly higher 6.8\,pc derived from estimates of \citet{Harris2002} from STIS data.  We also note that our and the \citet{Mieske2013} lower $\Psi$ values relative to \citet{Taylor2010} results are in part due to higher population $M/L$ estimates; \citet{Taylor2010} uses the \citet{BC2003} models, while \citet{Mieske2013} defines a somewhat higher $M/L$ vs. [Fe/H] relation that we also use in this paper.

From our findings here, in combination with the much lower $M/L$ found in M59cO by \citet{Ahn2017} relative to previous integrated-light studies, it is clear that at least some fraction of the integrated-light $M/L$s are not well determined. In general, there appears to be a bias towards overestimating the $M/L$.  This could be due to overestimation of the dispersion, perhaps due to galaxy light contaminating the integrated-light spectra of the UCD, errors in the light/mass profile determination of the UCDs, or modeling errors. Modeling of adaptive optics data with HST-based mass models is therefore key to assessing the reliability of previous integrated measurements.

         \begin{figure}
   \centering
   \includegraphics[width=\hsize]{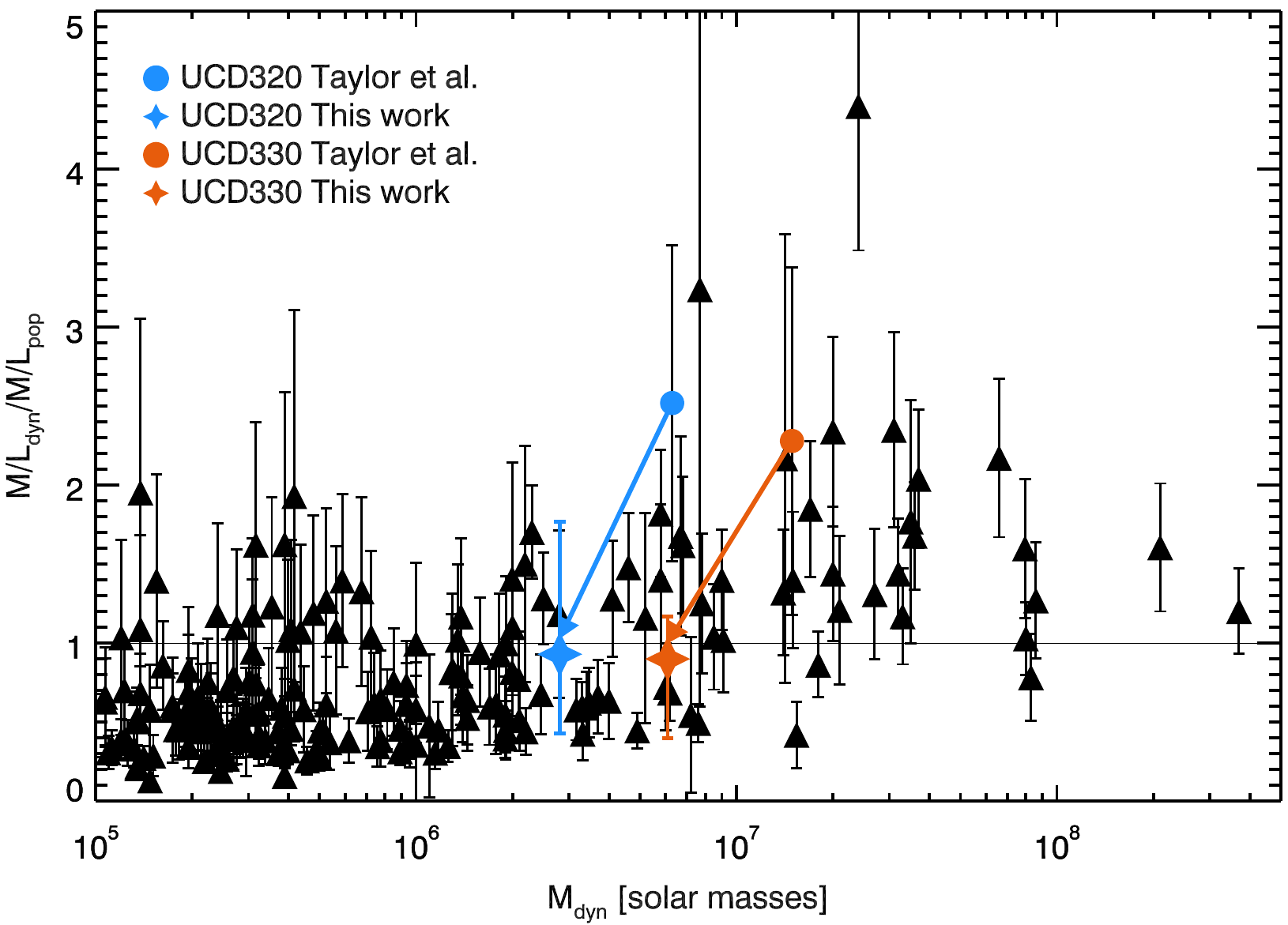}
      \caption{The ratio of $M/L_{\rm dyn}$ over the stellar population prediction $M/L_{\rm pop}$ is plotted versus the stellar mass of GCs and UCDs. The black datapoints are mainly from \cite{Mieske2013} with updated values for M60-UCD1 \citep{Seth2014}, M59cO and VUCD3 \citep{Ahn2017}, and M59-UCD3 (\citealt{Liu2015b}). The colored circles denote the measured values from \cite{Taylor2010} for both UCDs and the star symbols mark the new measured $\Psi$ values from this work, with an arrow marking the change towards the new measurements. The black line is where $M/L_{\rm dyn}$ is exactly equal to the expected stellar population $M/L_{\rm pop}$.}
         \label{fig:ML}
   \end{figure}

\subsection{Are UCD\,330 and 320 Stripped Nuclei?}

In this work, we have presented the first adaptive optics measurements of UCDs below $10^{7}M_{\odot}$.  Adaptive optics measurements of massive UCDs ($> 10^{7}M_{\odot}$) have found high-mass fraction BHs in all systems studied thus far \citep[][Afanasiev et al., {\em in prep}]{Seth2014,Ahn2017}.  These observations, combined with the higher than expected integrated-dispersion mass-to-light ratios in most of these systems \citep{Mieske2013}, suggests a large fraction of these UCDs have massive BHs. This strongly supports the model that many of these UCDs are stripped nuclei. Simulations of tidal stripping in cluster environments has shown that the predicted number of massive UCDs $\gtrsim 2 \times 10^{7}M_{\odot}$ are consistent with the observed numbers in Virgo and Fornax \citep{Pfeffer2014,Pfeffer2016}. At lower masses, a larger fraction of UCDs are likely to be ``ordinary'' globular clusters, not stripped nuclei, but this fraction is likely dependent on environment.

Any lower mass stripped nuclei probably originate from lower mass galaxies due to the scaling relation between NSCs and BHs \citep[e.g.][]{Ferrarese2006,Scott2013}. Assuming a UCD mass equal to the NSC mass, the compilation of \citet{Georgiev2016} suggests NSCs in the mass range of our objects ($2.5-6 \times10^{6}\,M_{\odot}$) have host galaxy stellar masses between $\sim10^{8}-5\times10^{10}\,M_{\odot}$.  The occupation fraction of BHs in this mass range of galaxies is not yet well known. Above $\sim10^9\,M_{\odot}$ the BH occupation fraction appears quite high in early-type galaxies \citep{Miller2015,Nguyen2017}, but local group galaxies M\,33 and NGC\,205 host no BHs above $\sim10^4M_{\odot}$ \citep{Gebhardt2001,Valluri2005}, while BHs of about this mass have been claimed in likely stripped nuclei M\,54 and $\omega$Cen and G1 \citep{Ibata2009,Noyola2010, Lutzgendorf2013}.  Our upper-limits in UCD\,320 and UCD\,330 do not exclude these intermediate-mass BHs, and thus our observations do not rule out that these objects could be the stripped nuclei of relatively low-mass galaxies $\lesssim 10^9$~M$_\odot$.

Another signature that may provide us clues about whether these objects are stripped nuclei is their metallicity. With [Fe/H]=$-$0.36 and $-$0.85 for UCD\,330 and UCD\,320 both objects have a significantly lower metallicity than any of the 4 high mass UCDs with BHs which all have at least solar metallicity. The high-metallicities for all UCDs above $\sim 3 \times 10^7$~M$_\odot$ was also suggested to be evidence for stripping amongst these systems by \citet{Janz2016}. Assuming that our systems are stripped nuclei, we can try to constrain their host masses based on their metallicites. 

The mass-metallicity relation of \citet{Lee2006} suggests galaxy masses of $\sim 10^9$~M$_\odot$ for UCD\,330 and $2 \times 10^7$~M$_\odot$ for UCD\,320.  However these are quite rough estimates, as the scatter in these relations is roughly an order of magnitude \citep{Tremonti2004}. \citet{Paudel2011} find nuclei with metallicities consistent with UCD\,320 in $\sim 10^9$~M$_\odot$ galaxies. For the case of UCD\,330, its mass and metallicity are fully consistent with being a stripped nucleus, while for UCD\,320, its NSC  would have to have been on the metal-poor end of the distribution given the mass of the NSC. However, their metallicities are also consistent with these objects being ordinary GCs.

We re-analyzed UVES data for UCD\,320 and UCD\,330 from \citet{Rejkuba2007} using the {\sc nbursts} stellar population fitting code as described in \citep{Chilingarian2007}. We find that the $\alpha$-abundance of UCD 330 is [$\alpha$/Fe]=+0.16$\pm$0.03\,dex, while that of UCD 320 is not well determined. The nuclei of dwarf galaxies are not significantly $\alpha$-enhanced \citep{Chilingarian2009,Paudel2011}, and thus our measurement of a moderate enhancement in UCD\,330 implies that it would be consistent with being the nucleus of a dwarf galaxy progenitor. The four UCDs with confirmed BHs all have higher enhancement in their $\alpha$-abundance between [$\alpha$/Fe]=0.2-0.5\,dex (\citealt{Francis2012}), indicating that they might originate from more massive progenitors, which have larger $\alpha$-abundance enhancements (\citealt{Thomas2005}).

Rotation may also be a signature of stripped galaxy nuclei. Strong rotation is seen in nearby nuclear star clusters \citep{Seth2008, Seth2010, Feldmeier2014, Nguyen2017}, with $V/\sigma$ values ranging between $\sim$0.3-1.3, with both early and late-type galaxies showing strong rotation. This rotation can be created through cluster merging, but the strongest rotation is likely to be due to {\em in situ} star formation \citep[e.g.][]{Hartmann2011,Tsatsi2017}.  Larger scale ($\sim$100\,pc) nuclear disks are also common \citep{Launhardt2002, Balcells2007, Chilingarian2009, Morelli2010, Toloba2014}, and stripping of these could also yield rotating UCDs. On the other hand, globular clusters typically do not rotate this strongly, with typical $V/\sigma \lesssim 0.2$ \citep{Lane2010, Bellazzini2012, Fabricius2014, Kimmig2015, Kamann2018}.  Given the measured $V/\sigma$ of UCD\,330 and 320 of 0.3 and 0.4 respectively, this relatively strong rotation may argue in favor of these objects being stripped nuclei. However, we caution that increasing rotation is also seen at longer relaxation times (\citealt{Kamann2018}), and our systems have longer relaxation times than most MW GCs (half light radius $t_{\rm rel}$ of 2.8\,Gyr for UCD\,330 and 6.0\,Gyr for UCD\,320).

Finally, many UCDs, including UCD\,330 have two component profiles, extratidal light, or tidal tails \citep{Martini2004,Evstigneeva2007, Wittmann2016, Voggel2016a}. These are expected for stripped galaxy nuclei, as the inner component tracks the original NSC, while the outer section is the remnants of the rest of the galaxy \citep{Pfeffer2013}.  Low surface-brightness halos found around galactic GCs have also been suggested to track tidal stripped nuclei \citep[e.g.][]{Olszewski2009,Kuzma2018}. Thus the two-component profile provides perhaps the strongest evidence for UCD\,330 being a stripped nucleus. 

In summary, the lack of massive BHs in UCD\,330 and 320 does not imply these systems aren't stripped nuclei. The relatively strong rotation and the two-component structure in UCD\,330 do support the idea that these UCDs may be stripped nuclei. In the case that they are stripped nuclei, they would probably come from low-mass galaxies where the BH demographics are not yet well understood.

\section{Conclusions}
\label{sec:conclusion}

This study is the first to target lower mass UCDs ($< 10^7$~M$_\odot$) in the search of central massive BHs, which would be strong evidence that they are the former nuclear star clusters of stripped dwarf galaxies. We constrain the BHs in these systems using dynamical modeling of stellar kinematic measurements from adaptive optics assisted VLT/SINFONI data.  We detect no BHs, however, we can place a $3\,\sigma$ upper limit on the BH masses. We find an upper limit of $M_{\bullet}=1.0\times10^{5}M_{\odot}$ for UCD\,330 and $M_{\bullet}=1.0\times10^{6}M_{\odot}$ for UCD\,320. This corresponds to relative mass fractions of 1.7\% and 37.7\% respectively, with the poorer constraint in UCD\,320 resulting from significantly worse data quality. The 1.7\% mass fraction upper limit in UCD\,330 excludes the presence of a high mass fraction (10-20\%) BH, similar to those found in more massive UCDs \citep[][Afanasiev et al., {\em in prep}]{Seth2014,Ahn2017}, however, an intermediate mass BH similar to those claimed in Local Group GCs \citep{Ibata2009,Noyola2010, Lutzgendorf2013} cannot be excluded.

We have shown that the dynamical $M/L$ of UCD\,320 and UCD\,330 are not inflated, and for both UCDS, our models are fully consistent with predictions from stellar population models with $\Psi=\frac{M_{\rm dyn}}{M_{\rm pop}}=0.9$ within the errorbars.

In most of the UCDs with measured massive BHs, the best-fit models without a BH suggest they are overmassive, with $\Psi > 1$ \citep{Mieske2013,Seth2014,Ahn2017}. Therefore, our BH non-detections in these low-mass UCDs supports the hypothesis that the inflated integrated-light dynamical $M/L$ found in many UCDs does indicate the presence of a high mass fraction BH.

Our study finds that both UCDs rotate significantly, which is often observed for nuclear star clusters, yet rarely for GCs.  Furthermore, the surface brightness profile of UCD\,330 is best-fit by a two component model, as expected for stripped nuclei. In UCD\,320, the high BH mass upper limit, combined with the X-ray source detected there still leaves room for this system to host a significant BH. Therefore, there is some support that these two UCDs may in fact be stripped dwarf galaxy nuclei.

With our upcoming program on SINFONI we will be able to test for the presence of a BH in three more low mass UCDs, more than doubling the sample of low-mass UCDs with resolved kinematics. We will be able to detect any potential BHs in them down to $2.0\times10^{5}M_{\odot}$. This will help to further establish whether stripped galaxy nuclei exist among low mass UCDs and to determine how the SMBH occupation fraction varies with UCD mass.

\acknowledgments

Work on this project by K.T. V.~and A.C.S.~was supported by AST-1350389. 
J.S. acknowledges support from NSF grant AST-1514763 and the Packard Foundation. A.J.R. was supported by National Science Foundation grant AST-1515084 and as a Research Corporation for Science Advancement Cottrell Scholar. IC's research is supported by the Telescope Data Center, Smithsonian Astrophysical Observatory. IC also acknowledges support by the Russian Science Foundation grant 17-72-20119 and by the ESO Visiting Scientist programme.
Based on observations made with the NASA/ESA Hubble Space Telescope, 
and obtained from the Hubble Legacy Archive, which is a collaboration 
between the Space Telescope Science Institute (STScI/NASA), the Space 
Telescope European Coordinating Facility (ST-ECF/ESA) and the 
Canadian Astronomy Data Centre (CADC/NRC/CSA).

%

\vspace{5mm}
\facilities{HST, VLT/SINFONI}





\bibliographystyle{apj}
\bibliography{bibliography_CenA}

\begin{thebibliography}{}
\expandafter\ifx\csname natexlab\endcsname\relax\def\natexlab#1{#1}\fi

\bibitem[{{Ahn} {et~al.}(2017){Ahn}, {Seth}, {den Brok}, {Strader},
  {Baumgardt}, {van den Bosch}, {Chilingarian}, {Frank}, {Hilker}, {McDermid},
  {Mieske}, {Romanowsky}, {Spitler}, {Brodie}, {Neumayer}, \&
  {Walsh}}]{Ahn2017}
{Ahn}, C.~P., {Seth}, A.~C., {den Brok}, M., {et~al.} 2017, \apj, 839, 72

\bibitem[{{Balcells} {et~al.}(2007){Balcells}, {Graham}, \&
  {Peletier}}]{Balcells2007}
{Balcells}, M., {Graham}, A.~W., \& {Peletier}, R.~F. 2007, \apj, 665, 1084

\bibitem[{{Bastian} {et~al.}(2006){Bastian}, {Saglia}, {Goudfrooij},
  {Kissler-Patig}, {Maraston}, {Schweizer}, \& {Zoccali}}]{Bastian2006}
{Bastian}, N., {Saglia}, R.~P., {Goudfrooij}, P., {et~al.} 2006, \aap, 448, 881

\bibitem[{{Baumgardt}(2017)}]{Baumgardt2017}
{Baumgardt}, H. 2017, \mnras, 464, 2174

\bibitem[{{Beasley} {et~al.}(2008){Beasley}, {Bridges}, {Peng}, {Harris},
  {Harris}, {Forbes}, \& {Mackie}}]{Beasley2008}
{Beasley}, M.~A., {Bridges}, T., {Peng}, E., {et~al.} 2008, \mnras, 386, 1443

\bibitem[{{Bekki} {et~al.}(2003){Bekki}, {Couch}, {Drinkwater}, \&
  {Shioya}}]{Bekki2003}
{Bekki}, K., {Couch}, W.~J., {Drinkwater}, M.~J., \& {Shioya}, Y. 2003, \mnras,
  344, 399

\bibitem[{{Bellazzini} {et~al.}(2012){Bellazzini}, {Dalessandro}, {Sollima}, \&
  {Ibata}}]{Bellazzini2012}
{Bellazzini}, M., {Dalessandro}, E., {Sollima}, A., \& {Ibata}, R. 2012,
  \mnras, 423, 844

\bibitem[{{Binney}(1978)}]{Binney1978}
{Binney}, J. 1978, \mnras, 183, 501

\bibitem[{{Brodie} {et~al.}(2011){Brodie}, {Romanowsky}, {Strader}, \&
  {Forbes}}]{Brodie2011}
{Brodie}, J.~P., {Romanowsky}, A.~J., {Strader}, J., \& {Forbes}, D.~A. 2011,
  \aj, 142, 199

\bibitem[{{Bruzual} \& {Charlot}(2003)}]{BC2003}
{Bruzual}, G., \& {Charlot}, S. 2003, \mnras, 344, 1000

\bibitem[{{Cappellari}(2002)}]{Cappellari2002}
{Cappellari}, M. 2002, \mnras, 333, 400

\bibitem[{{Cappellari}(2008)}]{Cappellari2008}
---. 2008, \mnras, 390, 71

\bibitem[{{Cappellari}(2017)}]{Cappellari2017}
---. 2017, \mnras, 466, 798

\bibitem[{{Cappellari} \& {Copin}(2003)}]{Cappellari2003}
{Cappellari}, M., \& {Copin}, Y. 2003, \mnras, 342, 345

\bibitem[{{Cappellari} \& {Emsellem}(2004)}]{Cappellari2004}
{Cappellari}, M., \& {Emsellem}, E. 2004, \pasp, 116, 138

\bibitem[{{Carretta} {et~al.}(2010){Carretta}, {Bragaglia}, {Gratton},
  {Lucatello}, {Bellazzini}, {Catanzaro}, {Leone}, {Momany}, {Piotto}, \&
  {D'Orazi}}]{Carretta2010}
{Carretta}, E., {Bragaglia}, A., {Gratton}, R.~G., {et~al.} 2010, \apjl, 714,
  L7

\bibitem[{{Chilingarian}(2009)}]{Chilingarian2009}
{Chilingarian}, I.~V. 2009, \mnras, 394, 1229

\bibitem[{{Chilingarian} {et~al.}(2007){Chilingarian}, {Prugniel},
  {Sil'Chenko}, \& {Afanasiev}}]{Chilingarian2007}
{Chilingarian}, I.~V., {Prugniel}, P., {Sil'Chenko}, O.~K., \& {Afanasiev},
  V.~L. 2007, \mnras, 376, 1033

\bibitem[{{Da Rocha} {et~al.}(2011){Da Rocha}, {Mieske}, {Georgiev}, {Hilker},
  {Ziegler}, \& {Mendes de Oliveira}}]{DaRocha2011}
{Da Rocha}, C., {Mieske}, S., {Georgiev}, I.~Y., {et~al.} 2011, \aap, 525, A86

\bibitem[{{Drinkwater} {et~al.}(2003){Drinkwater}, {Gregg}, {Hilker}, {Bekki},
  {Couch}, {Ferguson}, {Jones}, \& {Phillipps}}]{Drinkwater2003}
{Drinkwater}, M.~J., {Gregg}, M.~D., {Hilker}, M., {et~al.} 2003, \nat, 423,
  519

\bibitem[{{Drinkwater} {et~al.}(2000){Drinkwater}, {Jones}, {Gregg}, \&
  {Phillipps}}]{Drinkwater2000}
{Drinkwater}, M.~J., {Jones}, J.~B., {Gregg}, M.~D., \& {Phillipps}, S. 2000,
  \pasa, 17, 227

\bibitem[{{Eisenhauer} {et~al.}(2003){Eisenhauer}, {Abuter}, {Bickert},
  {Biancat-Marchet}, {Bonnet}, {Brynnel}, {Conzelmann}, {Delabre}, {Donaldson},
  {Farinato}, {Fedrigo}, {Genzel}, {Hubin}, {Iserlohe}, {Kasper},
  {Kissler-Patig}, {Monnet}, {Roehrle}, {Schreiber}, {Stroebele}, {Tecza},
  {Thatte}, \& {Weisz}}]{Eisenhauer2003}
{Eisenhauer}, F., {Abuter}, R., {Bickert}, K., {et~al.} 2003, in \procspie,
  Vol. 4841, Instrument Design and Performance for Optical/Infrared
  Ground-based Telescopes, ed. M.~{Iye} \& A.~F.~M. {Moorwood}, 1548--1561

\bibitem[{{Evstigneeva} {et~al.}(2007){Evstigneeva}, {Gregg}, {Drinkwater}, \&
  {Hilker}}]{Evstigneeva2007}
{Evstigneeva}, E.~A., {Gregg}, M.~D., {Drinkwater}, M.~J., \& {Hilker}, M.
  2007, \aj, 133, 1722

\bibitem[{{Fabricius} {et~al.}(2014){Fabricius}, {Noyola}, {Rukdee}, {Saglia},
  {Bender}, {Hopp}, {Thomas}, {Opitsch}, \& {Williams}}]{Fabricius2014}
{Fabricius}, M.~H., {Noyola}, E., {Rukdee}, S., {et~al.} 2014, \apjl, 787, L26

\bibitem[{{Feldmeier} {et~al.}(2014){Feldmeier}, {Neumayer}, {Seth},
  {Sch{\"o}del}, {L{\"u}tzgendorf}, {de Zeeuw}, {Kissler-Patig}, {Nishiyama},
  \& {Walcher}}]{Feldmeier2014}
{Feldmeier}, A., {Neumayer}, N., {Seth}, A., {et~al.} 2014, \aap, 570, A2

\bibitem[{{Ferrarese} {et~al.}(2006){Ferrarese}, {C{\^o}t{\'e}}, {Jord{\'a}n},
  {Peng}, {Blakeslee}, {Piatek}, {Mei}, {Merritt}, {Milosavljevi{\'c}},
  {Tonry}, \& {West}}]{Ferrarese2006}
{Ferrarese}, L., {C{\^o}t{\'e}}, P., {Jord{\'a}n}, A., {et~al.} 2006, \apjs,
  164, 334

\bibitem[{{Francis} {et~al.}(2012){Francis}, {Drinkwater}, {Chilingarian},
  {Bolt}, \& {Firth}}]{Francis2012}
{Francis}, K.~J., {Drinkwater}, M.~J., {Chilingarian}, I.~V., {Bolt}, A.~M., \&
  {Firth}, P. 2012, \mnras, 425, 325

\bibitem[{{Gebhardt} {et~al.}(2001){Gebhardt}, {Lauer}, {Kormendy}, {Pinkney},
  {Bower}, {Green}, {Gull}, {Hutchings}, {Kaiser}, {Nelson}, {Richstone}, \&
  {Weistrop}}]{Gebhardt2001}
{Gebhardt}, K., {Lauer}, T.~R., {Kormendy}, J., {et~al.} 2001, \aj, 122, 2469

\bibitem[{{Georgiev} {et~al.}(2016){Georgiev}, {B{\"o}ker}, {Leigh},
  {L{\"u}tzgendorf}, \& {Neumayer}}]{Georgiev2016}
{Georgiev}, I.~Y., {B{\"o}ker}, T., {Leigh}, N., {L{\"u}tzgendorf}, N., \&
  {Neumayer}, N. 2016, \mnras, 457, 2122

\bibitem[{{Girardi} {et~al.}(2000){Girardi}, {Bressan}, {Bertelli}, \&
  {Chiosi}}]{Padova2000}
{Girardi}, L., {Bressan}, A., {Bertelli}, G., \& {Chiosi}, C. 2000, \aaps, 141,
  371

\bibitem[{{Graham} \& {Spitler}(2009)}]{Graham2009}
{Graham}, A.~W., \& {Spitler}, L.~R. 2009, \mnras, 397, 2148

\bibitem[{{Harris} {et~al.}(1992){Harris}, {Geisler}, {Harris}, \&
  {Hesser}}]{Harris1992}
{Harris}, G.~L.~H., {Geisler}, D., {Harris}, H.~C., \& {Hesser}, J.~E. 1992,
  \aj, 104, 613

\bibitem[{{Harris} {et~al.}(2010){Harris}, {Rejkuba}, \& {Harris}}]{Harris2010}
{Harris}, G.~L.~H., {Rejkuba}, M., \& {Harris}, W.~E. 2010, \pasa, 27, 457

\bibitem[{{Harris} {et~al.}(2002){Harris}, {Harris}, {Holland}, \&
  {McLaughlin}}]{Harris2002}
{Harris}, W.~E., {Harris}, G.~L.~H., {Holland}, S.~T., \& {McLaughlin}, D.~E.
  2002, \aj, 124, 1435

\bibitem[{{Hartmann} {et~al.}(2011){Hartmann}, {Debattista}, {Seth},
  {Cappellari}, \& {Quinn}}]{Hartmann2011}
{Hartmann}, M., {Debattista}, V.~P., {Seth}, A., {Cappellari}, M., \& {Quinn},
  T.~R. 2011, \mnras, 418, 2697

\bibitem[{{Hernandez} {et~al.}(2018){Hernandez}, {Larsen}, {Trager}, {Kaper},
  \& {Groot}}]{Hernandez2018}
{Hernandez}, S., {Larsen}, S., {Trager}, S., {Kaper}, L., \& {Groot}, P. 2018,
  \mnras, arXiv:1802.09534

\bibitem[{{Hilker}(2006)}]{Hilker2006}
{Hilker}, M. 2006, ArXiv Astrophysics e-prints, astro-ph/0605447

\bibitem[{{Hilker} {et~al.}(1999){Hilker}, {Infante}, {Vieira},
  {Kissler-Patig}, \& {Richtler}}]{Hilker1999}
{Hilker}, M., {Infante}, L., {Vieira}, G., {Kissler-Patig}, M., \& {Richtler},
  T. 1999, \aaps, 134, 75

\bibitem[{{Hilker} {et~al.}(2004){Hilker}, {Kayser}, {Richtler}, \&
  {Willemsen}}]{Hilker2004}
{Hilker}, M., {Kayser}, A., {Richtler}, T., \& {Willemsen}, P. 2004, \aap, 422,
  L9

\bibitem[{{Ibata} {et~al.}(2009){Ibata}, {Bellazzini}, {Chapman},
  {Dalessandro}, {Ferraro}, {Irwin}, {Lanzoni}, {Lewis}, {Mackey}, {Miocchi},
  \& {Varghese}}]{Ibata2009}
{Ibata}, R., {Bellazzini}, M., {Chapman}, S.~C., {et~al.} 2009, \apjl, 699,
  L169

\bibitem[{{Irwin} {et~al.}(2016){Irwin}, {Maksym}, {Sivakoff}, {Romanowsky},
  {Lin}, {Speegle}, {Prado}, {Mildebrath}, {Strader}, {Liu}, \&
  {Miller}}]{Irwin2016}
{Irwin}, J.~A., {Maksym}, W.~P., {Sivakoff}, G.~R., {et~al.} 2016, \nat, 538,
  356

\bibitem[{{Janz} {et~al.}(2016){Janz}, {Norris}, {Forbes}, {Huxor},
  {Romanowsky}, {Frank}, {Escudero}, {Faifer}, {Forte}, {Kannappan},
  {Maraston}, {Brodie}, {Strader}, \& {Thompson}}]{Janz2016}
{Janz}, J., {Norris}, M.~A., {Forbes}, D.~A., {et~al.} 2016, \mnras, 456, 617

\bibitem[{{Jennings} {et~al.}(2015){Jennings}, {Romanowsky}, {Brodie}, {Janz},
  {Norris}, {Forbes}, {Martinez-Delgado}, {Fagioli}, \& {Penny}}]{Jennings2015}
{Jennings}, Z.~G., {Romanowsky}, A.~J., {Brodie}, J.~P., {et~al.} 2015, \apjl,
  812, L10

\bibitem[{{Kamann} {et~al.}(2018){Kamann}, {Husser}, {Dreizler}, {Emsellem},
  {Weilbacher}, {Martens}, {Bacon}, {den Brok}, {Giesers}, {Krajnovi{\'c}},
  {Roth}, {Wendt}, \& {Wisotzki}}]{Kamann2018}
{Kamann}, S., {Husser}, T.-O., {Dreizler}, S., {et~al.} 2018, \mnras, 473, 5591

\bibitem[{{Kimmig} {et~al.}(2015){Kimmig}, {Seth}, {Ivans}, {Strader},
  {Caldwell}, {Anderton}, \& {Gregersen}}]{Kimmig2015}
{Kimmig}, B., {Seth}, A., {Ivans}, I.~I., {et~al.} 2015, \aj, 149, 53

\bibitem[{{Kuzma} {et~al.}(2018){Kuzma}, {Da Costa}, \& {Mackey}}]{Kuzma2018}
{Kuzma}, P.~B., {Da Costa}, G.~S., \& {Mackey}, A.~D. 2018, \mnras, 473, 2881

\bibitem[{{Lane} {et~al.}(2010){Lane}, {Kiss}, {Lewis}, {Ibata}, {Siebert},
  {Bedding}, {Sz{\'e}kely}, {Balog}, \& {Szab{\'o}}}]{Lane2010}
{Lane}, R.~R., {Kiss}, L.~L., {Lewis}, G.~F., {et~al.} 2010, \mnras, 406, 2732

\bibitem[{{Launhardt} {et~al.}(2002){Launhardt}, {Zylka}, \&
  {Mezger}}]{Launhardt2002}
{Launhardt}, R., {Zylka}, R., \& {Mezger}, P.~G. 2002, \aap, 384, 112

\bibitem[{{Lee} {et~al.}(2006){Lee}, {Skillman}, {Cannon}, {Jackson}, {Gehrz},
  {Polomski}, \& {Woodward}}]{Lee2006}
{Lee}, H., {Skillman}, E.~D., {Cannon}, J.~M., {et~al.} 2006, \apj, 647, 970

\bibitem[{{Liu} {et~al.}(2015){Liu}, {Peng}, {Toloba}, {Mihos}, {Ferrarese},
  {Alamo-Mart{\'{\i}}nez}, {Zhang}, {C{\^o}t{\'e}}, {Cuillandre}, {Cunningham},
  {Guhathakurta}, {Gwyn}, {Herczeg}, {Lim}, {Puzia}, {Roediger},
  {S{\'a}nchez-Janssen}, \& {Yin}}]{Liu2015b}
{Liu}, C., {Peng}, E.~W., {Toloba}, E., {et~al.} 2015, \apjl, 812, L2

\bibitem[{{L{\"u}tzgendorf} {et~al.}(2013){L{\"u}tzgendorf}, {Kissler-Patig},
  {Neumayer}, {Baumgardt}, {Noyola}, {de Zeeuw}, {Gebhardt}, {Jalali}, \&
  {Feldmeier}}]{Lutzgendorf2013}
{L{\"u}tzgendorf}, N., {Kissler-Patig}, M., {Neumayer}, N., {et~al.} 2013,
  \aap, 555, A26

\bibitem[{{Maraston}(2005)}]{Maraston2005}
{Maraston}, C. 2005, \mnras, 362, 799

\bibitem[{{Maraston} {et~al.}(2004){Maraston}, {Bastian}, {Saglia},
  {Kissler-Patig}, {Schweizer}, \& {Goudfrooij}}]{Maraston2004}
{Maraston}, C., {Bastian}, N., {Saglia}, R.~P., {et~al.} 2004, \aap, 416, 467

\bibitem[{{Martini} \& {Ho}(2004)}]{Martini2004}
{Martini}, P., \& {Ho}, L.~C. 2004, \apj, 610, 233

\bibitem[{{Mieske} {et~al.}(2013){Mieske}, {Frank}, {Baumgardt},
  {L{\"u}tzgendorf}, {Neumayer}, \& {Hilker}}]{Mieske2013}
{Mieske}, S., {Frank}, M.~J., {Baumgardt}, H., {et~al.} 2013, \aap, 558, A14

\bibitem[{{Mieske} {et~al.}(2004){Mieske}, {Hilker}, \& {Infante}}]{Mieske2004}
{Mieske}, S., {Hilker}, M., \& {Infante}, L. 2004, \aap, 418, 445

\bibitem[{{Mieske} {et~al.}(2012){Mieske}, {Hilker}, \& {Misgeld}}]{Mieske2012}
{Mieske}, S., {Hilker}, M., \& {Misgeld}, I. 2012, \aap, 537, A3

\bibitem[{{Miller} {et~al.}(2015){Miller}, {Gallo}, {Greene}, {Kelly}, {Treu},
  {Woo}, \& {Baldassare}}]{Miller2015}
{Miller}, B.~P., {Gallo}, E., {Greene}, J.~E., {et~al.} 2015, \apj, 799, 98

\bibitem[{{Minniti} {et~al.}(1998){Minniti}, {Kissler-Patig}, {Goudfrooij}, \&
  {Meylan}}]{Minniti1998}
{Minniti}, D., {Kissler-Patig}, M., {Goudfrooij}, P., \& {Meylan}, G. 1998,
  \aj, 115, 121

\bibitem[{{Misgeld} \& {Hilker}(2011)}]{Misgeld2011}
{Misgeld}, I., \& {Hilker}, M. 2011, \mnras, 414, 3699

\bibitem[{{Morelli} {et~al.}(2010){Morelli}, {Cesetti}, {Corsini}, {Pizzella},
  {Dalla Bont{\`a}}, {Sarzi}, \& {Bertola}}]{Morelli2010}
{Morelli}, L., {Cesetti}, M., {Corsini}, E.~M., {et~al.} 2010, \aap, 518, A32

\bibitem[{{Murray}(2009)}]{Murray2009}
{Murray}, N. 2009, \apj, 691, 946

\bibitem[{{Nguyen} {et~al.}(2017){Nguyen}, {Seth}, {Neumayer}, {Kamann},
  {Voggel}, {Cappellari}, {Picotti}, {Nguyen}, {B{\"o}eker}, {Debattista},
  {Caldwell}, {McDermid}, \& {Nathan}}]{Nguyen2017}
{Nguyen}, D.~D., {Seth}, A.~C., {Neumayer}, N., {et~al.} 2017, ArXiv e-prints,
  arXiv:1711.04314

\bibitem[{{Norris} {et~al.}(2015){Norris}, {Escudero}, {Faifer}, {Kannappan},
  {Forte}, \& {van den Bosch}}]{Norris2015}
{Norris}, M.~A., {Escudero}, C.~G., {Faifer}, F.~R., {et~al.} 2015, ArXiv
  e-prints, arXiv:1506.00004

\bibitem[{{Norris} \& {Kannappan}(2011)}]{Norris2011}
{Norris}, M.~A., \& {Kannappan}, S.~J. 2011, \mnras, 414, 739

\bibitem[{{Norris} {et~al.}(2014){Norris}, {Kannappan}, {Forbes}, {Romanowsky},
  {Brodie}, {Faifer}, {Huxor}, {Maraston}, {Moffett}, {Penny}, {Pota},
  {Smith-Castelli}, {Strader}, {Bradley}, {Eckert}, {Fohring}, {McBride},
  {Stark}, \& {Vaduvescu}}]{Norris2014}
{Norris}, M.~A., {Kannappan}, S.~J., {Forbes}, D.~A., {et~al.} 2014, \mnras,
  443, 1151

\bibitem[{{Noyola} {et~al.}(2010){Noyola}, {Gebhardt}, {Kissler-Patig},
  {L{\"u}tzgendorf}, {Jalali}, {de Zeeuw}, \& {Baumgardt}}]{Noyola2010}
{Noyola}, E., {Gebhardt}, K., {Kissler-Patig}, M., {et~al.} 2010, \apjl, 719,
  L60

\bibitem[{{Olszewski} {et~al.}(2009){Olszewski}, {Saha}, {Knezek},
  {Subramaniam}, {de Boer}, \& {Seitzer}}]{Olszewski2009}
{Olszewski}, E.~W., {Saha}, A., {Knezek}, P., {et~al.} 2009, \aj, 138, 1570

\bibitem[{{Paudel} {et~al.}(2011){Paudel}, {Lisker}, \&
  {Kuntschner}}]{Paudel2011}
{Paudel}, S., {Lisker}, T., \& {Kuntschner}, H. 2011, \mnras, 413, 1764

\bibitem[{{Peng} {et~al.}(2002){Peng}, {Ho}, {Impey}, \& {Rix}}]{Peng2002}
{Peng}, C.~Y., {Ho}, L.~C., {Impey}, C.~D., \& {Rix}, H.-W. 2002, \aj, 124, 266

\bibitem[{{Pfeffer} \& {Baumgardt}(2013)}]{Pfeffer2013}
{Pfeffer}, J., \& {Baumgardt}, H. 2013, \mnras, 433, 1997

\bibitem[{{Pfeffer} {et~al.}(2014){Pfeffer}, {Griffen}, {Baumgardt}, \&
  {Hilker}}]{Pfeffer2014}
{Pfeffer}, J., {Griffen}, B.~F., {Baumgardt}, H., \& {Hilker}, M. 2014, \mnras,
  444, 3670

\bibitem[{{Pfeffer} {et~al.}(2016){Pfeffer}, {Hilker}, {Baumgardt}, \&
  {Griffen}}]{Pfeffer2016}
{Pfeffer}, J., {Hilker}, M., {Baumgardt}, H., \& {Griffen}, B.~F. 2016, \mnras,
  458, 2492

\bibitem[{{Rejkuba} {et~al.}(2007){Rejkuba}, {Dubath}, {Minniti}, \&
  {Meylan}}]{Rejkuba2007}
{Rejkuba}, M., {Dubath}, P., {Minniti}, D., \& {Meylan}, G. 2007, \aap, 469,
  147

\bibitem[{{Renaud} {et~al.}(2015){Renaud}, {Bournaud}, \& {Duc}}]{Renaud2015}
{Renaud}, F., {Bournaud}, F., \& {Duc}, P.-A. 2015, \mnras, 446, 2038

\bibitem[{{Rossa} {et~al.}(2006){Rossa}, {van der Marel}, {B{\"o}ker},
  {Gerssen}, {Ho}, {Rix}, {Shields}, \& {Walcher}}]{Rossa2006}
{Rossa}, J., {van der Marel}, R.~P., {B{\"o}ker}, T., {et~al.} 2006, \aj, 132,
  1074

\bibitem[{{Schulz} {et~al.}(2015){Schulz}, {Pflamm-Altenburg}, \&
  {Kroupa}}]{Schulz2015}
{Schulz}, C., {Pflamm-Altenburg}, J., \& {Kroupa}, P. 2015, \aap, 582, A93

\bibitem[{{Schweizer} {et~al.}(2018){Schweizer}, {Seitzer}, {Whitmore},
  {Kelson}, \& {Villanueva}}]{Schweizer2018}
{Schweizer}, F., {Seitzer}, P., {Whitmore}, B.~C., {Kelson}, D.~D., \&
  {Villanueva}, E.~V. 2018, \apj, 853, 54

\bibitem[{{Scott} \& {Graham}(2013)}]{Scott2013}
{Scott}, N., \& {Graham}, A.~W. 2013, \apj, 763, 76

\bibitem[{{Seth} {et~al.}(2008{\natexlab{a}}){Seth}, {Ag{\"u}eros}, {Lee}, \&
  {Basu-Zych}}]{Seth2008b}
{Seth}, A., {Ag{\"u}eros}, M., {Lee}, D., \& {Basu-Zych}, A.
  2008{\natexlab{a}}, \apj, 678, 116

\bibitem[{{Seth}(2010)}]{Seth2010}
{Seth}, A.~C. 2010, \apj, 725, 670

\bibitem[{{Seth} {et~al.}(2008{\natexlab{b}}){Seth}, {Blum}, {Bastian},
  {Caldwell}, \& {Debattista}}]{Seth2008}
{Seth}, A.~C., {Blum}, R.~D., {Bastian}, N., {Caldwell}, N., \& {Debattista},
  V.~P. 2008{\natexlab{b}}, \apj, 687, 997

\bibitem[{{Seth} {et~al.}(2006){Seth}, {Dalcanton}, {Hodge}, \&
  {Debattista}}]{Seth2006}
{Seth}, A.~C., {Dalcanton}, J.~J., {Hodge}, P.~W., \& {Debattista}, V.~P. 2006,
  \aj, 132, 2539

\bibitem[{{Seth} {et~al.}(2014){Seth}, {van den Bosch}, {Mieske}, {Baumgardt},
  {Brok}, {Strader}, {Neumayer}, {Chilingarian}, {Hilker}, {McDermid},
  {Spitler}, {Brodie}, {Frank}, \& {Walsh}}]{Seth2014}
{Seth}, A.~C., {van den Bosch}, R., {Mieske}, S., {et~al.} 2014, \nat, 513, 398

\bibitem[{{She} {et~al.}(2017){She}, {Ho}, \& {Feng}}]{She2017}
{She}, R., {Ho}, L.~C., \& {Feng}, H. 2017, \apj, 842, 131

\bibitem[{{Siegel} {et~al.}(2007){Siegel}, {Dotter}, {Majewski}, {Sarajedini},
  {Chaboyer}, {Nidever}, {Anderson}, {Mar{\'{\i}}n-Franch}, {Rosenberg},
  {Bedin}, {Aparicio}, {King}, {Piotto}, \& {Reid}}]{Siegel2007}
{Siegel}, M.~H., {Dotter}, A., {Majewski}, S.~R., {et~al.} 2007, \apjl, 667,
  L57

\bibitem[{{Taylor} {et~al.}(2010){Taylor}, {Puzia}, {Harris}, {Harris},
  {Kissler-Patig}, \& {Hilker}}]{Taylor2010}
{Taylor}, M.~A., {Puzia}, T.~H., {Harris}, G.~L., {et~al.} 2010, \apj, 712,
  1191

\bibitem[{{Thomas} {et~al.}(2005){Thomas}, {Maraston}, {Bender}, \& {Mendes de
  Oliveira}}]{Thomas2005}
{Thomas}, D., {Maraston}, C., {Bender}, R., \& {Mendes de Oliveira}, C. 2005,
  \apj, 621, 673

\bibitem[{{Toloba} {et~al.}(2014){Toloba}, {Guhathakurta}, {van de Ven},
  {Boissier}, {Boselli}, {den Brok}, {Falc{\'o}n-Barroso}, {Hensler}, {Janz},
  {Laurikainen}, {Lisker}, {Paudel}, {Peletier}, {Ry{\'s}}, \&
  {Salo}}]{Toloba2014}
{Toloba}, E., {Guhathakurta}, P., {van de Ven}, G., {et~al.} 2014, \apj, 783,
  120

\bibitem[{{Tremonti} {et~al.}(2004){Tremonti}, {Heckman}, {Kauffmann},
  {Brinchmann}, {Charlot}, {White}, {Seibert}, {Peng}, {Schlegel}, {Uomoto},
  {Fukugita}, \& {Brinkmann}}]{Tremonti2004}
{Tremonti}, C.~A., {Heckman}, T.~M., {Kauffmann}, G., {et~al.} 2004, \apj, 613,
  898

\bibitem[{{Tsatsi} {et~al.}(2017){Tsatsi}, {Mastrobuono-Battisti}, {van de
  Ven}, {Perets}, {Bianchini}, \& {Neumayer}}]{Tsatsi2017}
{Tsatsi}, A., {Mastrobuono-Battisti}, A., {van de Ven}, G., {et~al.} 2017,
  \mnras, 464, 3720

\bibitem[{{Valluri} {et~al.}(2005){Valluri}, {Ferrarese}, {Merritt}, \&
  {Joseph}}]{Valluri2005}
{Valluri}, M., {Ferrarese}, L., {Merritt}, D., \& {Joseph}, C.~L. 2005, \apj,
  628, 137

\bibitem[{{van der Marel} \& {Anderson}(2010)}]{vanderMarel2010}
{van der Marel}, R.~P., \& {Anderson}, J. 2010, \apj, 710, 1063

\bibitem[{{Voggel} {et~al.}(2016){Voggel}, {Hilker}, \&
  {Richtler}}]{Voggel2016a}
{Voggel}, K., {Hilker}, M., \& {Richtler}, T. 2016, \aap, 586, A102

\bibitem[{{Walcher} {et~al.}(2006){Walcher}, {B{\"o}ker}, {Charlot}, {Ho},
  {Rix}, {Rossa}, {Shields}, \& {van der Marel}}]{Walcher2006}
{Walcher}, C.~J., {B{\"o}ker}, T., {Charlot}, S., {et~al.} 2006, \apj, 649, 692

\bibitem[{{Wallace} \& {Hinkle}(1996)}]{Wallace1996}
{Wallace}, L., \& {Hinkle}, K. 1996, \apjs, 107, 312

\bibitem[{{Wittmann} {et~al.}(2016){Wittmann}, {Lisker}, {Pasquali}, {Hilker},
  \& {Grebel}}]{Wittmann2016}
{Wittmann}, C., {Lisker}, T., {Pasquali}, A., {Hilker}, M., \& {Grebel}, E.~K.
  2016, \mnras, 459, 4450

\bibitem[{{Zocchi} {et~al.}(2017){Zocchi}, {Gieles}, \&
  {H{\'e}nault-Brunet}}]{Zocchi2017}
{Zocchi}, A., {Gieles}, M., \& {H{\'e}nault-Brunet}, V. 2017, \mnras, 468, 4429

\end{thebibliography}

\appendix

\begin{deluxetable}{cccc}
\tablecaption{Multi-Gaussian Expansion of UCD\,330 that provides the luminosity model for the JAM code. The horizontal line separates the two components of the S\'ersic model.\label{tab:MGE330}}
\tablehead{\colhead{Luminosity} & \colhead{$\sigma$} & \colhead{$q$} & \colhead{Position Angle} \\ 
\colhead{$\frac{L_{\odot}}{\rm pc^{2}}$} & \colhead{$\arcsec$} & \colhead{} &  \colhead{$^{\circ}$}} 

\startdata
   50415.66 &     0.0012	&     0.841  &   $-$48.84  \\ 
   71147.24  &    0.0040  &    0.841   &   $-$48.84  \\ 
   81974.73  &    0.01078  &    0.841  &    $-$48.84  \\ 
   74298.20  &  0.0251   &     0.841   &   $-$48.84  \\ 
   50359.54  &    0.0516  &     0.841  &    $-$48.84  \\ 
   24333.01  &   0.0962    &     0.841  &    $-$48.84  \\ 
    7999.0797  &    0.1651   &    0.841   &   $-$48.84  \\ 
    1718.68  &     0.2651    &     0.841 &     $-$48.84  \\ 
     229.46  &   0.4079 &    0.841   &   $-$48.84  \\ 
     12.88  &    0.6372    &    0.841   &   $-$48.84  \\ 
      \hline
  244219.16   &   0.0004   &  0.8000    &   $-$48.69  \\ 
  174900.37   &   0.0010    &  0.8000    &   $-$48.69  \\ 
  117329.34 &    0.0025     &  0.8000    &   $-$48.69  \\ 
   72208.21  &    0.0061     &  0.8000    &   $-$48.69  \\ 
   39796.08  &   0.0140       &  0.8000    &   $-$48.69  \\ 
   19754.61  &   0.0309    &  0.8000    &   $-$48.69  \\ 
   8705.59   &     0.0657   &  0.8000    &   $-$48.69  \\ 
    3384.10   &  0.1350     &  0.8000    &   $-$48.69  \\ 
     1151.80  &    0.2690   &  0.8000    &   $-$48.69  \\ 
     337.71  &     0.5191     &  0.8000    &   $-$48.69  \\ 
      87.01  &    0.9691     &  0.8000    &   $-$48.69  \\ 
      19.30  &    1.7639   &  0.8000    &   $-$48.69  \\ 
       3.63  &    3.1265   &  0.8000    &   $-$48.69  \\ 
       0.59  &     5.4029    &  0.8000    &   $-$48.69  \\ 
       0.08  &   9.1426    &  0.8000    &   $-$48.69  \\ 
       0.01  &   15.4174      &  0.8000    &   $-$48.69  \\ 
       0.001  &  28.2834    &  0.8000    &   $-$48.69  \\ 
\enddata

\end{deluxetable}

\begin{deluxetable}{cccc}
\tablecaption{Multi-Gaussian Expansion of UCD\,320 that provides the luminosity model for the JAM code\label{tab:MGE320}}
\tablehead{\colhead{Luminosity} & \colhead{$\sigma$} & \colhead{$q$} & \colhead{Position Angle} \\ 
\colhead{$\frac{L_{\odot}}{\rm pc^{2}}$} & \colhead{$\arcsec$} & \colhead{} &  \colhead{$^{\circ}$}} 
\startdata
297084.18  &     0.0005  &         0.646  &      $-$79.79  \\
  256401.63  &     0.0012    &      0.646  &     $-$79.79  \\
  200800.78  &    0.0031  &      0.646  &      $-$79.79  \\
  140971.31  &     0.0073  &        0.646  &     $-$79.79  \\
   88486.19  &      0.0162    &      0.646  &       $-$79.79  \\
   48435.90  &   0.0346    &       0.646  &       $-$79.79  \\
   22142.39   &      0.0705   &       0.646  &     $-$79.79  \\
    8633.37  &     0.1367  &       0.646  &        $-$79.79  \\
    2894.15  &    0.2541   &      0.646  &      $-$79.79  \\
     795.18  &     0.4570   &      0.646  &     $-$79.79  \\
     179.44  &     0.7942   &      0.646  &      $-$79.79  \\
      32.91  &    1.3367   &       0.646  &       $-$79.79  \\
       4.95  &     2.1895  &        0.646  &     $-$79.79  \\
       0.58 &   3.5684 &       0.646  &        $-$79.79  \\
       0.04  &   6.2344   &        0.646  &       $-$79.79  \\
\enddata

\end{deluxetable}



\end{document}